\documentclass[manuscript]{aastex}

\shorttitle{Dust Environment of Comet 22P/Kopff}
\shortauthors{Moreno et al.}

\begin{document}


\title{Comet 22P/Kopff: Dust environment and grain ejection anisotropy from visible and infrared observations}


\author{Fernando Moreno\affil{Instituto de Astrof\'\i sica de Andaluc\'\i a, CSIC,
  Glorieta de la Astronom\'\i a s/n, 18008 Granada, Spain}
\email{fernando@iaa.es}}

\author{
Francisco Pozuelos\affil{Instituto de Astrof\'\i sica 
de Andaluc\'\i a, CSIC, Glorieta de la Astronom\'\i a s/n, 
18008 Granada, Spain} }

\author{
Francisco Aceituno\affil{Instituto de Astrof\'\i sica 
de Andaluc\'\i a, CSIC, Glorieta de la Astronom\'\i a s/n, 
18008 Granada, Spain} }

\author{
V\'\i ctor Casanova\affil{Instituto de Astrof\'\i sica 
de Andaluc\'\i a, CSIC, Glorieta de la Astronom\'\i a s/n, 
18008 Granada, Spain} }

\author{
Alfredo Sota\affil{Instituto de Astrof\'\i sica 
de Andaluc\'\i a, CSIC, Glorieta de la Astronom\'\i a s/n, 
18008 Granada, Spain} }

\author{
Julio Castellano\affil{Amateur Association Cometas-Obs, Spain} }

\and

\author{
Esteban Reina\affil{Amateur Association Cometas-Obs, Spain} }




\begin{abstract}

We present optical observations and Monte Carlo models of the dust coma, tail, and trail structures of 
comet 22P/Kopff during the 2002 and 2009 apparitions. Dust loss rates, ejection velocities, and power-law  
size distribution functions are derived as functions of the 
heliocentric distance using pre- and post-perihelion imaging observations during 
both apparitions. The 2009 post-perihelion images can 
be accurately fitted by an isotropic ejection model.  On the other hand, 
strong dust ejection anisotropies are required to fit the near-coma regions at large heliocentric distances (both inbound at $r_h$=2.5 AU and outbound at 
$r_h$=2.6 AU) for the 2002 apparition. These asymmetries are compatible with a scenario where 
dust ejection is mostly seasonally-driven, coming mainly from regions near subsolar latitudes at far heliocentric distances inbound and outbound. At intermediate to near-perihelion heliocentric distances, the outgassing would affect much more extended latitude regions, the emission becoming almost isotropic near perihelion. We derived a maximum dust production rate of 260 kg s$^{-1}$ at perihelion, and an averaged production rate over one orbit of 40 kg s$^{-1}$. An enhanced emission rate, accompanied also by a large ejection velocity, is predicted at $r_h>$2.5 pre-perihelion. 

The model has also been extended to the thermal infrared in order to be applied to available trail observations with IRAS and ISO spacecrafts of this comet. The modeled trail intensities are in good agreement with those observations, which is remarkable taking into account that those data are sensitive to dust ejection patterns corresponding to several orbits before the 2002 and 2009 apparitions. 

\end{abstract}




\section{Introduction}

Comet 22P/Kopff is a Jupiter-family comet with a current orbital period 
of 6.43 years and perihelion distance of $q$=1.577 Astronomical Units (AU). 
In March 1954, it approached Jupiter at a distance of only 
0.17 AU, which induced a shortening in perihelion distance and 
orbital period. \cite{Yeomans74} found significant non-gravitational effects 
for this object, with secular changes of quasi-regular nature in the 
parameter $A_2$. These changes have been attributed to nucleus 
precession \citep{Yeomans74,Sekanina84,Rickman87}. Nucleus size
determinations converge to a value near $R_N$=1.8 km \citep{Lamy02,Tancredi00, Lowry01, Groussin09},  while the geometric albedo is estimated at $p_v$=0.042$\pm$0.006 \citep{Lamy02}. 

Estimates of the dust production rate for comet 22P are based on
a limited amount of data. \cite{Lamy02} reported 130 kg s$^{-1}$ near perihelion, at $r_h$=1.59 AU, an 
estimation based on the value of the measured Af$\rho$ parameter 
\citep{Ahearn84}. \cite{Ishiguro07} gave (710$\pm$70)($r_h$/AU)$^{-4}$ kg s$^{-1}$, based on optical 
trail images obtained during the 2002 apparition, roughly 
in accordance with the \cite{Lamy02} value at  $r_h$=1.59 AU, and an averaged production rate over one orbital period of 17$\pm$3 
kg s$^{-1}$. However, 
\cite{Sykes92} using trail images from IRAS satellite data reported  
$dM/dt$=3.16$\times$10$^{13}$ g century$^{-1}$, or 10 kg s$^{-1}$ averaged over one revolution. 


In order to report an accurate characterization of the comet dust environment, an extended set of observations covering a large fraction of the comet's orbit is needed. In this paper we combine post-perihelion image observations of 22P/Kopff obtained during the last 2009 apparition with pre- and post-perihelion archived images from the previous revolution around the Sun, the 2002 apparition. In addition, CCD lightcurves and Af$\rho$ data from amateur observers (the astronomical association {\it{Cometas-Obs}}), corresponding also to the 2002 and 2009 apparition, have been taken into account. We used our Monte Carlo dust tail modeling procedure in an attempt to fit the complete image set, which allows us to derive the dust parameters: size distribution, ejection velocities, mass loss rate, and ejection morphology. 

Once a best-fil model was found, we also considered the available trail data in the infrared, as reported by \cite{Sykes92} and \cite{Davies97}, from observations by IRAS and ISO spacecrafts. To that end, we developed a version of our Monte Carlo code to retrieve both optical and infrared fluxes.

\section{Observations}

We acquired images of the comet through a Johnson's $R$ bandpass using a 1024$\times$1024 CCD camera at Sierra Nevada Observatory (OSN) in Granada, Spain, at several epochs after the 2009 perihelion (2009-May-25.218). The pixel size on the sky was 0.46\arcsec, and the field of view 7.8\arcmin$\times$7.8\arcmin. Table I shows the log of the observations. The individual images at each night were bias subtracted and flat-fielded. At each night, the comet was repeatedly imaged and we combined the individual frames using a median stacking method. Whenever possible, calibration stars were also imaged. For all the other nights, the calibration was performed using field stars with the USNO-B1.0 star catalog \citep{Monet03}  which provides a 0.3 mag accuracy. The resulting images at each night were calibrated to mag arcsec$^{-2}$, and then converted to solar disk intensity units. In order to compare with the modeled images, and to improve the signal-to-noise ratio, we needed to rebin some of the images. Their final resolution at each date is indicated in Table I. 

For our modeling purposes, we also considered images from the previous comet orbit, i.e., images from the 2002 apparition. Specifically, we considered coma/trail images obtained at large heliocentric distances by Masateru Ishiguro at Kiso 1.05-m Schmidt telescope in Nagano, Japan, and Canada-France-Hawaii 3.6-m telescope (CFHT), previously described by \cite{Ishiguro07}. M. Ishiguro made kindly available to us the Kiso data, while the CFHT data were downloaded from the CFHT archive server. In both cases, we followed a similar reduction procedure as that described for the OSN data. We used field stars as calibration sources. Table I also lists the relevant information corresponding to those data sets.

Figure 1 displays all the final product images described above. As stated, some of them were rebinned in order to make them tractable with the Monte Carlo models and to improve the signal-to-noise ratio. All the images have been rotated to the $(N,M)$ photographic plane \citep{Finson68}, so that the Sun is towards the bottom. No contours have been displayed for the 2002/2003 images \citep{Ishiguro07} so that the neck-line/trail structures are visible. The neck-line, first described by \cite{Kimura77}, corresponds to large particles ejected at a true anomaly of 180$^\circ$ before the observation date, and appear to groundbased observers as a bright linear feature when the cometocentric latitude of the Earth is small (the Earth is near the comet orbital plane). A trail is formed by large particles that are ejected at low velocities from the nucleus so that they remain along the cometary orbit for many orbital periods. Neck-line structures are formed in the current orbital revolution, while trails correspond to particles ejected during the previous several orbits. In Figure 1, neck-line structures are clearly seen in e.g. panels (d) and (e), while the linear feature seen in (h) is clearly a trail, as this image was taken at large heliocentric distance ($r_h$=2.5 AU) pre-perihelion \citep{Ishiguro07}. 

In addition to the image data described above, we have also benefited from amateur observations carried out by the astronomical association {\it Cometas-Obs} (see http://astrosurf.com/Cometas-Obs), from both the 2002 and 2009 apparitions, providing a CCD lightcurve 
and Af$\rho$ measurements as a function of heliocentric distance. Both the magnitude and Af$\rho$ measurements are referred to an aperture of radius 10$^4$ km projected on the sky at each observation date. This choice is made to permit a direct comparison with the OSN Af$\rho$ data with at the same $\rho$ value. The data reduction was accomplished by using field stars and the CMC-14/USNO A2.0 star catalogs. Figure 2 presents the CCD lightcurve where the magnitudes $m$ have been reduced to heliocentric and geocentric distances of 1 AU by the equation: 

\begin{equation}
m(1,1,\alpha)=m-2.5\log(\Delta r_h^2)
\end{equation} 
  
where $\Delta$ and $r_h$ are the geocentric and heliocentric distances of the comet in AU, respectively. Most of the data points correspond to the 2009 apparition, where the  lightcurve shows a conspicuous maximum approximately 100 days from perihelion. In figure 2 the dependence of comet phase angle with time is also displayed. It is interesting to note that the maximum brightness corresponds to the minimum value of phase angle reached ($\alpha\sim$3.4$^\circ$). This means that the spike can be in principle attributed to the brightness opposition effect, although, overimposed to that, an enhancement of cometary activity during those dates post-perihelion cannot be ruled out. In order to show the brightness opposition effect, Figure 3 displays $m(1,1,\alpha)$ versus phase angle for those points having $r_h<$2 AU. As can be seen, these data can be well fitted by a linear phase coefficient of 0.028$\pm$0.002 mag deg$^{-1}$. This estimate is in the range of values obtained for other comets \citep[e.g.][]{Meech87}.

The maximum at 100 days post-perihelion also appears in the Af$\rho$ data, as could be expected. Figure 4 shows the Af$\rho$ data for a projected distance of $\rho$=10$^4$ km together with the Af$\rho$ data derived from the 2009 OSN images. The agreement between the {\it Cometas-Obs} group data and the OSN data for the 2009 apparition is excellent.  This emphasizes the importance of amateur astronomy groups in deriving information that is very useful for professional astronomers. Also  displayed are the Kiso/CFHT data and those from {\it Cometas-Obs} from the 2002 apparition. 
    
\section{The Model}

In order to model the images, we used a direct Monte Carlo 
dust tail code, which
is based on previous works of cometary dust tail analysis 
\citep[e.g.,][]{Moreno09,Moreno11}, and in the characterization of
the dust environment of comet 67P/Churyumov-Gerasimenko 
before Rosetta's arrival in 2014 \citep[the so-called Granada model, see][]{Fulle10}. Briefly, the code computes the trajectories of a
large number of particles ejected from a cometary nucleus surface, submitted 
to the solar gravity and radiation pressure fields. The gravity
of the comet itself is neglected, which is a good approximation for
22P owing to its small size (see section 1). The particles are  
accelerated to their terminal velocities 
by the gas molecules coming from the sublimating ices. These  
terminal velocities correspond to the input ejection  
velocities considered in the model. The particles describe a Keplerian
trajectory around the Sun, whose orbital elements are computed from
the terminal velocity and the $\beta$ parameter, which is the ratio of the 
force exerted by the solar radiation pressure and the solar gravity
\citep{Fulle89}. This parameter can be expressed as $\beta =
C_{pr}Q_{pr}/(2\rho_d r)$, where $C_{pr}$=1.19$\times$ 10$^{-3}$ kg
m$^{-2}$, and $\rho_d$ is the particle density, assumed $\rho_d$=1000 
kg m$^{-3}$ throughout. We used Mie theory for 
spherical particles to compute both the radiation pressure coefficient, 
$Q_{pr}$, and the geometric albedo, $p_v$. $Q_{pr}$ is a function of the  
particle radius, and $p_v$  is a function of the phase angle $\alpha$ and 
the particle radius. $p_v$ is  
obtained as $S_{11}(\alpha)\pi/k^2G$, where $S_{11}$ is the (1,1) element of the
scattering matrix, $k$=2$\pi$/$\lambda$ (the wavenumber), and $G$ is the 
geometrical cross section of the particle, i.e. $G$=$\pi r^2$ . We 
assumed the particles as glassy carbon spheres 
of refractive index $m$=1.88+0.71$i$, which is the value 
reported by \cite{Edoh83} for an incident wavelength of $\lambda$=0.6 $\mu$m. 
Highly absorbing 
particles have been often invoked to represent cometary material \citep[e.g.][]{Kimura03}. Figure 5 gives the dependence of $Q_{pr}$ and $p_v$ 
on particle radius. The values graphed for $p_v$ correspond to a phase angle of $\alpha$=10$^\circ$, but the results are identical for phase angles $\alpha <$40$^\circ$ for particle radii $r >$ 1 $\mu$m.  The asymptotic 
value of $p_v$ at  $r >$ 1 $\mu$m     
becomes $p_v$=0.036, which is within the error bar of the 
value of $p_v$=0.042$\pm$0.006 estimated by \cite{Lamy02}. Regarding 
$Q_{pr}$, we obtain $Q_{pr} \sim$1 for particle radii 
$r \stackrel{>}{\sim}$1 $\mu$m. The main difficulty when dealing with absorbing spherical particles is that the brightness opposition effect cannot be modeled, as the phase function keeps constant for phase angles $\alpha <$40$^\circ$. In principle, the scattering properties of non-spherical  particles can be computed with dedicated light scattering codes such as the Discrete Dipole Approximation or the T-matrix method \citep[e.g.,][]{Draine00, Mishchenko00}. However, the amount of CPU time and memory needed to make such calculations for realistic particle sizes becomes prohibitive, so that computations are only available for particle sizes of the order of the wavelength of the incident light or slightly larger \citep[e.g.,][]{Kimura03,  Kolokolova04, Moreno07, Zubko11}. 

For each observation date, the trajectories
of a large number of dust particles are computed, and their
positions on the $(N,M)$ plane are calculated. Then, their contribution to
the tail brightness is computed. The synthetic tail 
brightness obtained at each date depends on the
ejection velocity law assumed, the particle size
distribution, the dust mass loss rate, and the geometric albedo of the
particles, apart from the ejection pattern (anisotropy). 

As stated in the previous section, the image taken in 2002-05-12 \citep{Ishiguro07} shows a trail, which corresponds to dust ejected during many previous apparitions. The encounter of the comet with Jupiter in March 1954, modified its orbital elements \citep{Kepinski58,Kepinski63} in such a way that the particles ejected prior to 1954 probably followed very distinct orbits to those ejected since then. In consequence, for the interpretation of trails we considered that their age is the difference between the observation date and March 1954. We also assumed that the dust ejection pattern did not change with time: we do not have any information on dust ejection parameters back to 1954, so that there are no reasons to assume any temporal change. 

\section{Results and discussion}

In addition to the assumptions described in the previous section 
on some of the parameters involved in the 
model, and owing to their large number, we needed to 
make more additional hypotheses. Thus, the particle velocity is 
parametrized as
$v(t,\beta)=v_1(t)\beta^{1/2}$, where $v_1(t)$ is a time-dependent function to be determined in the modeling. In principle, we
assumed that $v_1(t)$ was a symmetric function of the heliocentric distance. 
Regarding the dust mass loss rate, we assumed an time-dependent asymmetric function with respect to perihelion based on the lightcurve described in section 2. Excluding the brightness opposition effect, for heliocentric distances $r_h>$2 AU, the light curve is clearly asymmetric, indicating higher production rates pre- than post-perihelion. 

The particle size distribution is assumed to be described by a power 
law with a constant power index, independent of the heliocentric 
distance. The minimum and maximum values for the particle radii must also be determined. In principle, there are no constraints for the minimum particle size. However, the maximum size ejected is constrained by the escape velocity, $v_{esc}=(2GM_c/R_{cm})^{1/2}$, where $G$ is the universal gravitational constant, $M_c$ is the comet mass, and $R_{cm}$ is the distance to the nucleus center of mass. For comet 22P, the mass has been determined from non-gravitational forces by \cite{Sosa09}, who reported  $M_c$=5.3$\times$10$^{12}$ kg. We assume that for a distance of $\sim$20 nuclear radii the effect of outgassing has vanished so that for $R_N$=1.8 km as given above, we get $v_{esc}$=18.8 cm s$^{-1}$.

 \subsection{Isotropic ejection models}

We first assumed the most simple model to fit the observations, i.e.,  
an isotropic ejection model. With the 
set of assumptions described above, we attempted first to fit all the observations together, i.e., those OSN images   
corresponding to the 2009 and those of the 2002 apparitions \citep{Ishiguro07}, assuming the same model inputs.  Since it is not practical to show the modeling results for each parameter combination, we will give just a summary of how we proceeded to find the final model parameters.   
Our first modeling attempts 
adopting the maximum particle sizes as constrained by the escape velocity, which implies ejection of particles in excess of $r$=14 cm near perihelion, resulted in very bright neck-line structures in each image, much brighter than observed. In addition, the trail intensities for the modeled 20020512 image were also too high. In consequence, we had to decrease that upper limit by an order of magnitude, to 1.4 cm. With this upper limit, we started to find models which produced a closer approximation to the full dataset.  Also, we realized that both symmetric dust mass loss rates and ejection velocities did not give acceptable fits.  Much better fits were found for asymmetric curves, with a steeper decay of activity post-perihelion than pre-perihelion. This is expected, based on the asymmetry observed in both the magnitude and the Af$\rho$ parameter.  The best fits are obtained for a size distribution function having a power index of $\alpha$=--3.1 and a minimum particle size of 1 $\mu$m, both parameters being independent of the heliocentric distance. Figure 6 gives the dependence of dust mass loss rates and velocities on heliocentric distance. The comparison of the isophote fields for all the images is given in figure 7. As it is seen the agreement between the observed and modeled images for the 2009 apparition is quite good (see Figs. 7a to 7g), except for the 20090828 image (Figure 7c) where the phase angle is minimum ($\alpha$=3.9$^\circ$), and the brightness opposition effect is maximum. As stated above, this effect cannot be modeled 
by using absorbing spherical particles, as the phase function does not experience any increase towards backscattering, so that the synthetic image remains darker than observed. The maximum dust loss rate corresponds to perihelion, with a value of 260 kg s$^{-1}$. This is twice the value reported by \cite{Lamy02}, who gave 130 kg s$^{-1}$ near perihelion, although their estimation corresponds to the earlier 1996 apparition. 

When the same isotropic model is applied to the images of the previous orbital revolution, several fitting problems were encountered (see Figs. 7h to 7k). Regarding the 20030731 image, while the overall observed and modeled intensity levels agree, the experimental isophotes depart significantly from the modeled ones, which are very asymmetric, as already noted by \cite{Ishiguro07}. In the case of the 20020512 image, there is a mismatch not only in the overall intensities, but also in the shape of the isophotes, that disagree strongly, as well in the trail intensities, which are much lower than observed (Figure 8). The coma/tail region of the 20020512 synthetic image is far weaker than observed, implying a higher dust mass loss rate prior to the observations. In fact, the Af$\rho$ measurements by {\it{Cometas-Obs}} group (figure 4) confirm a high activity of 22P at those heliocentric distances pre-perihelion ($r_h >$2.5 AU), with Af$\rho$ values comparable to those found near-perihelion.  In an attempt to fit the intensity levels of the 20020512 image, we increased both the mass loss rate and the particle ejection velocities at distances $r_h >$2.5 pre-perihelion. We found that the mass loss rate of figure 6 should be multiplied by a factor of order 10 at least, and the ejection velocity by a factor of 5, at $r_h>$2.5. The effect of doing these modifications is shown in Figure 9. In this figure, the mass loss rates are increased by a factor of 15, and the velocity by a factor of 5, at $r_h>$2.5 AU. As it is seen, there is now a better match between isophotes for the 20020512 image, although a higher factor would give a still better agreement (see next subsection). On the other hand, these changes improve the fits on all the 2009 images, as the neck-line structures appear broader and closer to the observations than those with the reduced mass loss rate (compare, e.g. figures 7 and 9, panels (d) to (g)), which is a consequence of the larger velocities imposed at $r_h >$2.5 AU pre-perihelion. 

We then focused on the asymmetries found in the isophote field in the near-nucleus region, on these two images of the 2002 apparition. We will show in the next subsection that the analysis of those images provides  clues on the dust ejection anisotropy.

 \subsection{Anisotropic ejection models} 

As explained in the previous subsection, while the 2009 apparition post-perihelion images can be adequately matched with isotropic ejection models, the 20020512, and 20030731 images from the previous orbit show clear departures from an isotropic dust ejection scenario. In order to fit these images, we then tried anisotropic ejection models. In principle, we experimented with a rotating spherical nucleus with an active area on it, with a rotation period of 12.3 hours \citep{Lowry03}. The spin axis orientation is being specified by two angles, the argument of the subsolar meridian at perihelion, $\Phi$, and the obliquity $I$ \citep{Sekanina81}. The active area is specified by setting a latitude-longitude box where the particles are assume to be ejected from. Since the age of the tails is much longer than a rotational period, the choice of a particular longitude range turns out to be irrelevant for the resulting synthetic images.

The approach consisted in considering as inputs the best-fit parameters found
for the isotropic ejection models, and then search for the best-fit $I$ and $\Phi$ and latitude box [$\lambda_{min}$,$\lambda_{max}$], for each of the two images, 20020512, and 20030731. The minimization procedure was accomplished by using the downhill simplex method of \cite{Nelder65} with the FORTRAN implementation by \cite{Press92}. The function to minimize was the standard deviation of the synthetic image from the observation. We are aware that this procedure only gives a local minimum: the code was run repeatedly with a varied starting simplex so that a more generic solution could be found. 
The input mass loss rates and velocities were those assumed in the previous subsection, but with a higher factor for the mass loss rate of 30 at $r_h>$2.5 AU pre-perihelion, which produces a better fit to the 20020512 image. With this modified input mass, the total dust mass ejected per orbit becomes  8$\times$10$^9$ kg, with an averaged production rate per orbital revolution of 40 kg s$^{-1}$, or 1.3$\times$10$^9$ kg yr$^{-1}$, which represents a modest contribution to the interplanetary dust cloud as compared to the  1.6-6.3$\times$10$^{10}$ kg  yr$^{-1}$ of comet 29P/Schwassmann-Wachmann \citep{Moreno09}. In comparison,  \cite{Ishiguro07} obtained an orbital averaged contribution of only 17$\pm$3 kg s$^{-1}$ for 22P/Kopff, although these authors stated that since their data only cover a small portion of the comet orbit, they probably underestimated the production rate near perihelion. On the other hand, as already indicated in the introduction, \cite{Sykes92} obtained 10 kg s$^{-1}$, averaged over an orbital period, from IRAS satellite images. The origin of this discrepance is probably related to the fact that the trail data refer only to the large particle component of the dust ejection.

The best solutions for both the 20020512 and 20030731 images corresponded to $\Phi$ between 170$^\circ$ and 210$^\circ$, while $I$ was between 50$^\circ$  and 70$^\circ$, with latitude boxes of [-90$^\circ$ ,-70$^\circ$] for the 20020512 image and [70$^\circ$ ,90$^\circ$] for the 20030731 image. The values for which the overall best fits for the two images were $\Phi$=180$^\circ$, and $I$=60$^\circ$, so that the rotational axis points to either RA=3$^\circ$, DEC=25$^\circ$ (prograde), or the opposite direction for retrograde rotation, RA=183$^\circ$, DEC=--25$^\circ$, since the sense of rotation is unconstrained. This procedure served to find the location of the active areas. However, the resulting synthetic images had too sharp borders, with a very fast decrease of brightness outwards in some directions. In order to smooth out the resulting images, we implemented two procedures. One was to impose some fraction of particles being ejected isotropically. The other was to consider that every point source on the surface in the Monte Carlo procedure was actually an emission cone with a certain width $\Delta\phi$, so that each particle ejected from a given latitude, instead of being ejected normal to the spherical nucleus surface, is ejected with a random azimuthal angle around that direction and a random angle (smaller than or equal to the cone width) with respect to that normal direction. In this way, the sharp borders dissapear and the synthetic images become closer to the observations. Two additional parameters have then to be defined, the fraction of particles ejected isotropically, and the cone angle width, that can be, in principle, different for the two images under analysis (20020512 and 20030731). After many simulations, we concluded that for the 20020512 image, the best fit parameters were $\Delta\phi$=60$^\circ$, with no need for an isotropic ejection fraction. In contrast, for the 20030731 image, the best-fit parameters were $\Delta\phi$=20$^\circ$, with 30\% of particles being ejected isotropically. Figure 10 show the fits to the near-coma regions of these two images. While we recognize the complexity of these models, with such a large number of parameters, we also remark that they are actually needed in order to reproduce the details seen in the images with a certain degree of accuracy. As it is seen in figure 10, these synthetic images reproduce the observations very closely.  

The location of the inferred active areas for the 20020512 and 20030731 images differ drastically: while the southernmost latitude region is active prior to the 20050512 observation, it is the northermost latitude area the one which is most active prior ($r_h >$1.7 AU) to the 20030731 observation. In order to see whether or not there is some relation to solar insolation, figure 11  shows the location of the subsolar point as a function of the heliocentric distance. As can be seen, the northern high latitudes ($\lambda\sim$50$^\circ$) are indeed those exposed to the highest solar radiation at heliocentric distances $r_h>$2.5 AU pre-perihelion, and, conversely, the southern high latitudes ($\lambda\sim$--50$^\circ$)  are those exposed to highest radiation after perihelion. Therefore, although the active areas we have inferred are somewhat displaced towards the north pole for the 20020512 simulation, and towards the south pole for the 20030731 simulation, they are close to the subsolar point. Therefore, these 2002/2003 data are consistent with the picture in which the region near the subsolar point is the first to experience significant outgassing inbound, while it is the latest to be active outbound. 

The final step was to consider a model that is able to fit all the available observations. We have attempted to fit such asymmetric ejection model to the whole image set, including both the 2003 and 2009 apparition images. In order to fit the 20020512 and 20030731 images, we have just shown that the active area at heliocentric distances $r_h>$2.5 pre-perihelion should be located near the north pole, and near the south pole at heliocentric distances $r_h>$1.7 post-perihelion, in such a way that the subsolar point is close to those areas. On the other hand, we have shown previously that all the 2009 post-perihelion images can be fitted by assuming isotropic ejection.  Therefore, the logical approach would be to assume that the active area limits converge towards the narrow latitude regions found for the two images 20020512 and 20030731 when the comet is at large heliocentric distance inbound and outbound, respectively. Those borders would define a broader latitude region at intermediate heliocentric distances. Figure 12 shows the upper and lower latitude limits of the active area as a function of the heliocentric distance, in which we have also drawn the location of the subsolar point. As we can see, the location on the active area correlates with the location of the subsolar point. Also, as the derived cone widths were different for the 20020512 and 20030731 images, we also set this parameter as time-dependent. Since the best fits were $\Delta\phi$=60$^\circ$ at $r_h>$2.5 AU pre-perihelion, and $\Delta\phi$=20$^\circ$ at $r_h\stackrel{>}{\sim}$1.7 post-perihelion, we just considered an interpolated solution in between, as shown in Figure 12 and also in Table II, which summarizes all the parameters that apply to the final version of the model. In addition, a fraction of 30\% of particles were assumed to undergo isotropic ejection, for all heliocentric distances. Figure 13 shows the results of the synthetic isophotes compared with the observations. It is clear that the asymmetric model reproduces with great detail most features observed at each epoch for both 2002 and 2009 apparitions. 

Finally, the 20020512 synthetic trail intensity is compared with the observation in figure 14. The agreement is also remarkable. This asymmetric ejection model has then proved to be valid for all the observations covering a large fraction of the comet orbit, and two consecutive apparitions.

 \section{Application of the dust model to observations of trails in the infrared} 

In order to take into account all available observations of dust from comet 22P/Kopff, we also considered the application of the final version of the anisotropic dust model to trail observations in the infrared, such as the IRAS data reported by \cite{Sykes92}, and the ISO/IRAS data by \cite{Davies97}. To do that, we developed a modified version of our Monte Carlo code to retrieve thermal fluxes. The thermal radiation from a single grain is given by:

\begin{equation}
F_\lambda={r^2\over\Delta^2}\epsilon(\lambda,r)\pi B_\lambda[T(r)]
\end{equation}

where $\epsilon(\lambda,r)$ is the grain emissivity at wavelength $\lambda$, and $B(\lambda,T)$ is the Planck function for grain temperature $T$ \citep{Hanner97}. The emissivities are computed by applying Kirchoff's law, so that $\epsilon(\lambda,r)$=$Q_{abs}(\lambda,r)$, where $Q_{abs}(\lambda,r)$ is the absorption efficiency of the grain. This quantity is computed by Mie theory for glassy carbon spheres, the same composition assumed to compute the scattering parameters at red wavelengths, using the variation of refractive index with wavelength reported by \cite{Edoh83}. 
   
The temperature from a single grain at a given heliocentric distance is  
computed from the balance between the energy absorbed in the visual and the
energy emitted in the infrared as \citep{Hanner97}:

\begin{equation}
{\pi r^2 \over r_h^2}\int_{0}^\infty S(\lambda) Q_{abs}(\lambda,r)d\lambda = 
4\pi r^2 \int_{0}^\infty Q_{abs}(\lambda,r)\pi B_\lambda[T(r)]d\lambda
\end{equation}

where $S(\lambda)$ is the solar flux at 1 AU, which we consider as a blackbody at $T$=5900 K. This equation is solved for $T(r)$ by Brent's algorithm. In this way, we generate a table from which we derive the equilibrium temperature as a function of particle radius and heliocentric distance by a two-dimensional interpolation. Then, using equation (2), we compute the infrared flux of a given particle as a function of the position in the $(N,M)$ plane. The rest of the Monte Carlo procedure is analogous as the one developed for the analysis at red wavelengths shown in previous sections. 

\cite{Davies97} obtained IR fluxes from 22P/Kopff trail images on March 26, 1992, using ISOCAM, the infrared camera of ISO satellite. They reported IR fluxes at 12 $\mu$m and trail widths at two positions of the trail behind the comet, one at +0.5$^\circ$ in mean anomaly, and another at +1$^\circ$ in mean anomaly. We have performed a simulation of the trail brightness at that wavelength, and at that date, as shown in figure 15. In that simulation we used a trail age of 38 years, which is the time spanned between the 1954 close encounter of the comet with Jupiter, and the observation. \cite{Davies97} also reported updated IRAS fluxes and widths of the dust trail of the comet at the same wavelength, and same positions in mean anomaly, but with the comet at a different heliocentric and geocentric distances. Since the IRAS images are built up of scans taken over a period of several weeks, the observation date selected in our code should be representative of the mean date of that period. For the IRAS observations, in Table II of \cite{Davies97}, there appear heliocentric distances of 1.69 and 1.66 AU, and geocentric distances of 1.48 and 1.45 AU, for the observations at mean anomalies of +0.5$^\circ$ and +1.0$^\circ$, respectively. These distances are approximately those that apply on October 15, 1983 ($r_h$=1.7 AU, $\Delta$=1.47 AU), which is the date we selected to compare with those IRAS observations. Therefore the trail age is 29.8 years. Table III shows the model results compared to IRAS and ISOCAM observations. As can be seen, there is a good agreement with the ISOCAM intensities, and with the IRAS intensity at $\delta MA$=+0.5$^\circ$, but not with the intensity at $\delta MA$=+1.0$^\circ$, where the modeled intensity is more than four times smaller than measured. Observations of 22P/Kopff trail were also performed with instrumentation on board the {\it Midcourse Space Experiment} mission \citep{Kraemer05}. The comet was observed about two months post-perihelion, on September 11, 1996, and the resulting trail maximum brightness at 12.1 $\mu$m wavelength and $\delta MA$=+0.16$^\circ$ was 0.74 MJy sr$^{-1}$, which is also significantly smaller than the IRAS result.    

On the other hand, the modeled trail widths are within the estimated errors of the IRAS measurements, but narrower than the ISOCAM widths in a factor 2-3. Since those trails contain information on the dust ejected  several orbits before the observations, these widths discrepancies could be well attributed to differences in ejection velocities over many orbits back to the 2009 apparition, to which the model actually applies. Nevertheless, the excellent agreement of the modeled intensities with the observations (with the exception of the IRAS intensity at  $\delta MA$=+1.0$^\circ$) is remarkable.

\section{Conclusions}
   
The analysis of a large image dataset of comet 22P/Kopff during two consecutive apparitions have permitted to develop a model by which the most relevant cometary dust parameters have been retrieved accurately. The observations have been compared to amateur astronomical observations by the {\it{Cometas-Obs}} group, showing a remarkable agreement in the behavior of  the Af$\rho$ parameter as a function of the heliocentric distance. Both the CCD lightcurve and Af$\rho$ observations indicate a brightness excess at 100 days from perihelion which is clearly correlated with the phase angle, and therefore indicative of a brightness opposition effect. The linear phase  coefficient is 0.028$\pm$0.002 mag deg$^{-1}$.    

 Assuming spherical particles of density of $\rho_d$=1000 kg m$^{-3}$, and glassy carbon composition (refractive index at red wavelengths of $m$=1.88+0.71$i$), which gives a geometric albedo of $p_v$=0.036,  the mass loss rate peaks near perihelion with a value of 260 kg s$^{-1}$. The comet onset of activity occurs at heliocentric distances $r_h\stackrel{>}{\sim}$3.5 AU, showing a clear time-dependent asymmetric behavior, with an enhanced activity at heliocentric distances beyond 2.5 AU pre-perihelion, accompanied also by an enhanced particle ejection velocity. The total mass ejected per orbit is 8$\times$10$^9$ kg, with an averaged dust mass loss rate per orbital period of 40 kg s$^{-1}$. The images can all be modeled assuming a time-constant differential size distribution function characterized by a power-law of index --3.1. 

The analysis of archive images corresponding to the 2002 apparition, acquired at large heliocentric distances pre- and post-perihelion provide clues on the dust ejection anisotropy. Such a model suggests an ejection scenario where the outgassing comes from regions near subsolar latitudes at far heliocentric distances pre- and post-perihelion, but becoming nearly isotropic at intermediate heliocentric distances pre- and near-perihelion. The anisotropic model is characterized by a latitude-dependent active area on a rotating nucleus with rotational parameters $\Phi$=180$^\circ$, $I$=60$^\circ$. The dust ejection is produced 
from emission cones randomly distributed on the active area surface with a time-dependent cone angle widths varying between 20$^\circ$ (at far distances pre-perihelion) and 60$^\circ$ (at far distances 
post-perihelion).    

We have demonstrated that such model is compatible with all the available observations. Besides, the trail intensities that characterize the time-averaged behavior of the large particle component of the dust are reproduced accurately, providing additional strength on the results obtained.       
 
In order to compare with infrared trail intensities obtained by IRAS and ISO satellites, a modified version of the Monte Carlo code has been developed to produce synthetic thermal images. The modeled intensities are in agreement with ISOCAM observations at $\delta MA$=+0.5$^\circ$, and at $\delta MA$=+1.0$^\circ$, while the trail widths are significantly narrower than reported by \cite{Davies97}. However, they are similar to the IRAS data \citep{Sykes92} when re-analyzed by \cite{Davies97}. Since those observations correspond to several orbits before the 2002 and 2009 apparitions to which the model actually applies, the discrepancies could just reflect real variations in dust ejection velocities and/or dust loss masses among different cometary orbits.  

\acknowledgments

We thank an anomymous referee for his/her comments and suggestions for improving the paper.  

We are grateful to M. Ishiguro, who provided granted access to Kiso Observatory 105-cm Schmidt telescope images. 

We are indebted to the amateur astronomers of the association {\it Cometas-Obs} for providing us with 22P/Kopff observations, in particular to   
I. Almendros, R. Naves, M. Campas, J.R. Vidal, F. Baldris, J.L. Salto, E.  Cort\'es, F. Garc\'\i a, J. Camarasa, J. Lopesino, J.M. Bosch, C. Gonz\'alez, 
C. Colazo, F. Tifner, D. Carde\~nosa, M. Rodr\'\i guez de Viguri, 
J.F. Hern\'andez, J.C. Mill\'an, F. Garc\'\i a, J.M. Ruiz, F.A. Rodr\'\i guez, C. Piret, G. Muler, J.A. Henr\'\i quez, O. Canales, R. Benavides, and J. Temprano.  

This research was based on data obtained at the Observatorio
de Sierra Nevada, which is operated by the Instituto de
Astrof\'\i sica de Andaluc\'\i a, CSIC.

Part of the data used in this paper were downloaded from the CFHT Science Data Archive. This research used the facilities of the Canadian Astronomy Data Centre operated by the National Research Council of Canada with the support of the Canadian Space Agency.   

This work was supported by contracts AYA2009-08011, and P09-FQM-4555 (Proyecto de Excelencia, Junta de Andaluc\i\' a).



{\it Facilities:} \facility{OSN}, \facility{Kiso}, \facility{CFHT}.



\clearpage

\begin{figure}
\includegraphics[angle=-90,scale=.75]{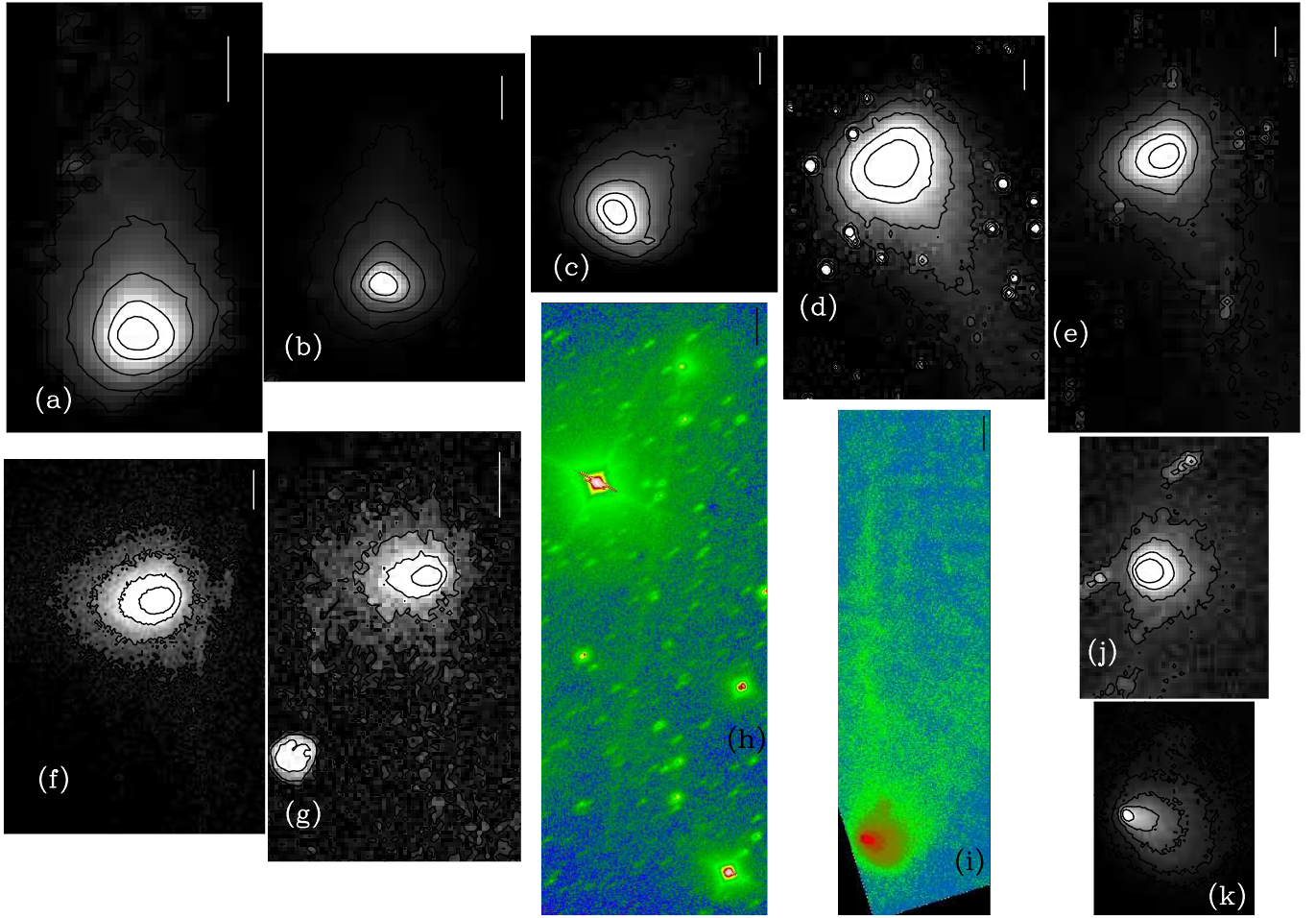}
\caption{Observations of comet 22P/Kopff through red filter bandpasses. Images (a) to (g) correspond to the 2009 apparition, and were obtained using a CCD camera at the 1.52-m telescope of the Observatorio de Sierra Nevada in Granada, Spain. The observation date of each image is as follows: (a), 2009-07-31; (b), 2009-08-15; (c), 2009-08-28; (d), 2009-09-21; (e), 2009-10-12; (f), 2009-11-09; (g), 2009-11-24. Panel (h) corresponds to Kiso observatory, acquired on 2002-05-12, and panel (i) corresponds to the CFHT observation on 2003-07-31 \citep{Ishiguro07}. Panels (j) and (k) are zoomed regions of images (h) and (i) near the coma regions. In all panels, vertical bars correspond to 20000 km on the sky, except for panel (h), where the bar represents 80000 km. See Table I for the associated physical parameters of each image.}
\end{figure}

\clearpage

\begin{figure}
\includegraphics[angle=-90,scale=.80]{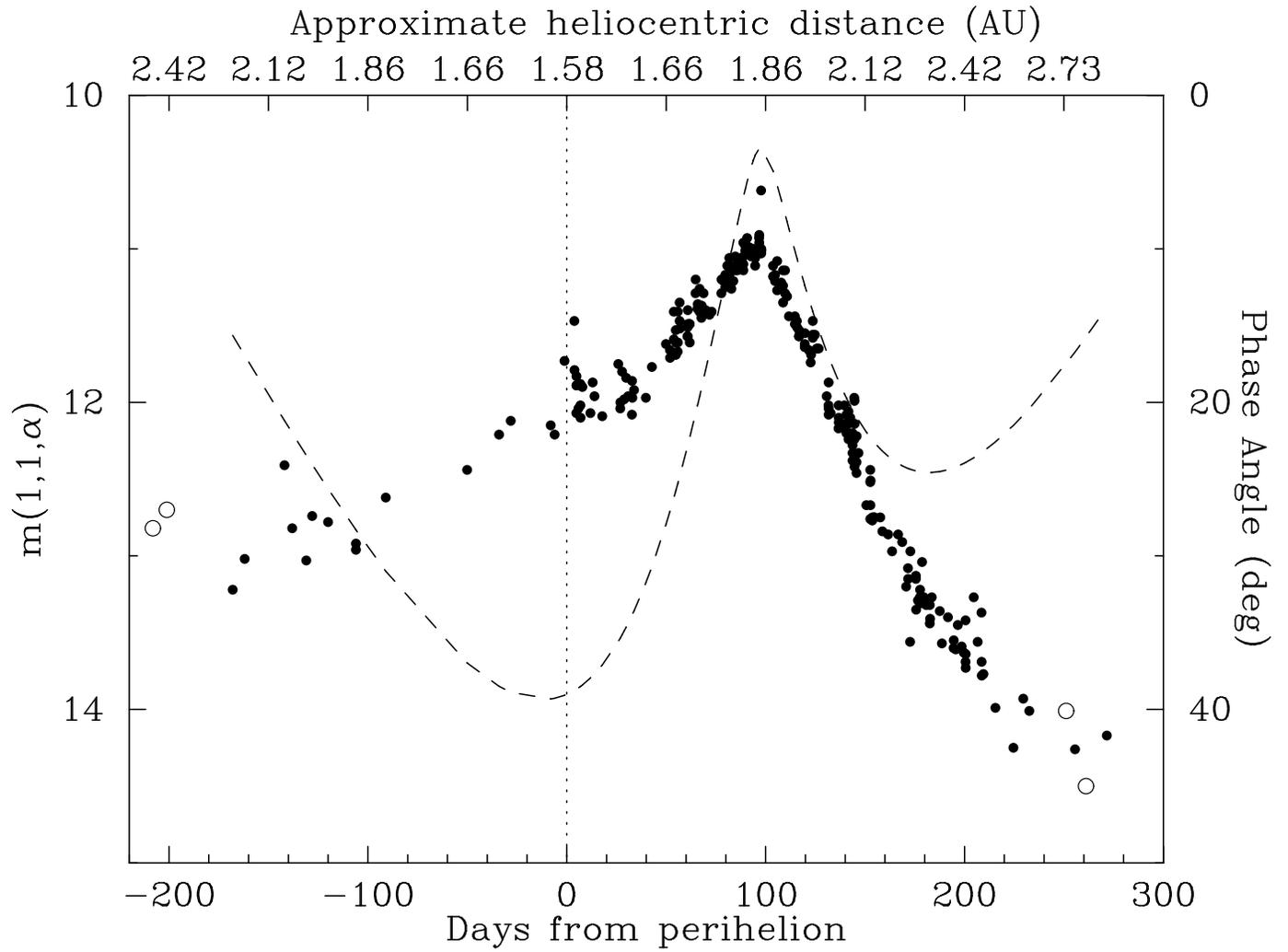}
\caption{Cometary magnitude $m(1,1,\alpha)$ as a function of time and heliocentric distance by {\it{Cometas-Obs}}. Filled circles correspond to the 2009 apparition and open circles to the 2002 apparition. The dashed line indicates the phase angle as a function of time.} 
\end{figure}
\clearpage

\begin{figure}
\includegraphics[angle=-90,scale=.80]{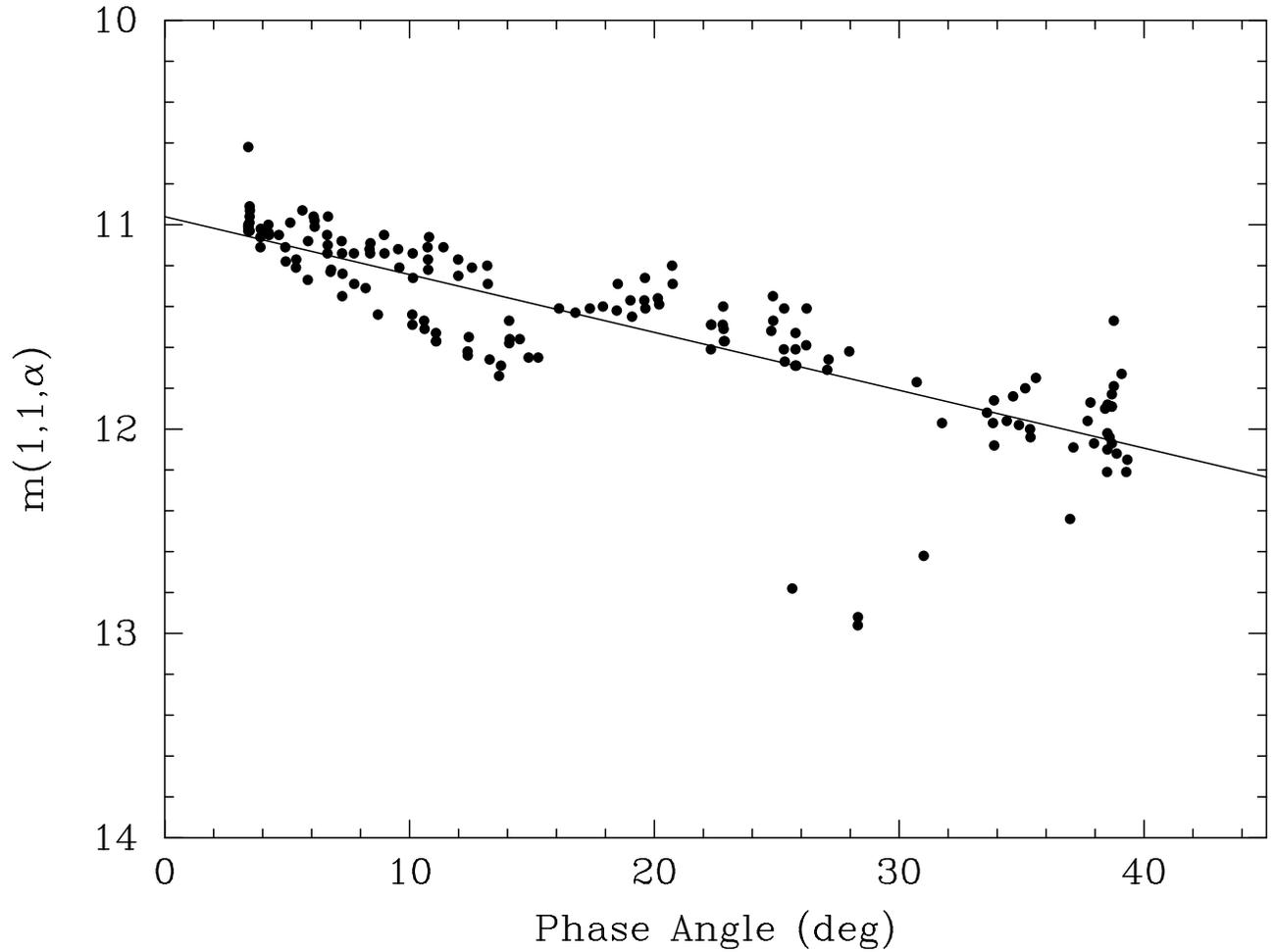}
\caption{Cometary magnitude $m(1,1,\alpha)$ as a function of phase angle for data points of $r_h<$2 AU, for the 2009 apparition. Filled circles are data from {\it Cometas-Obs}. The solid line represents a linear fit to the data,  whose slope corresponds to the linear phase coefficient, 0.028$\pm$0.002 mag deg$^{-1}$.}
\end{figure}
\clearpage

\begin{figure}
\includegraphics[angle=-90,scale=.80]{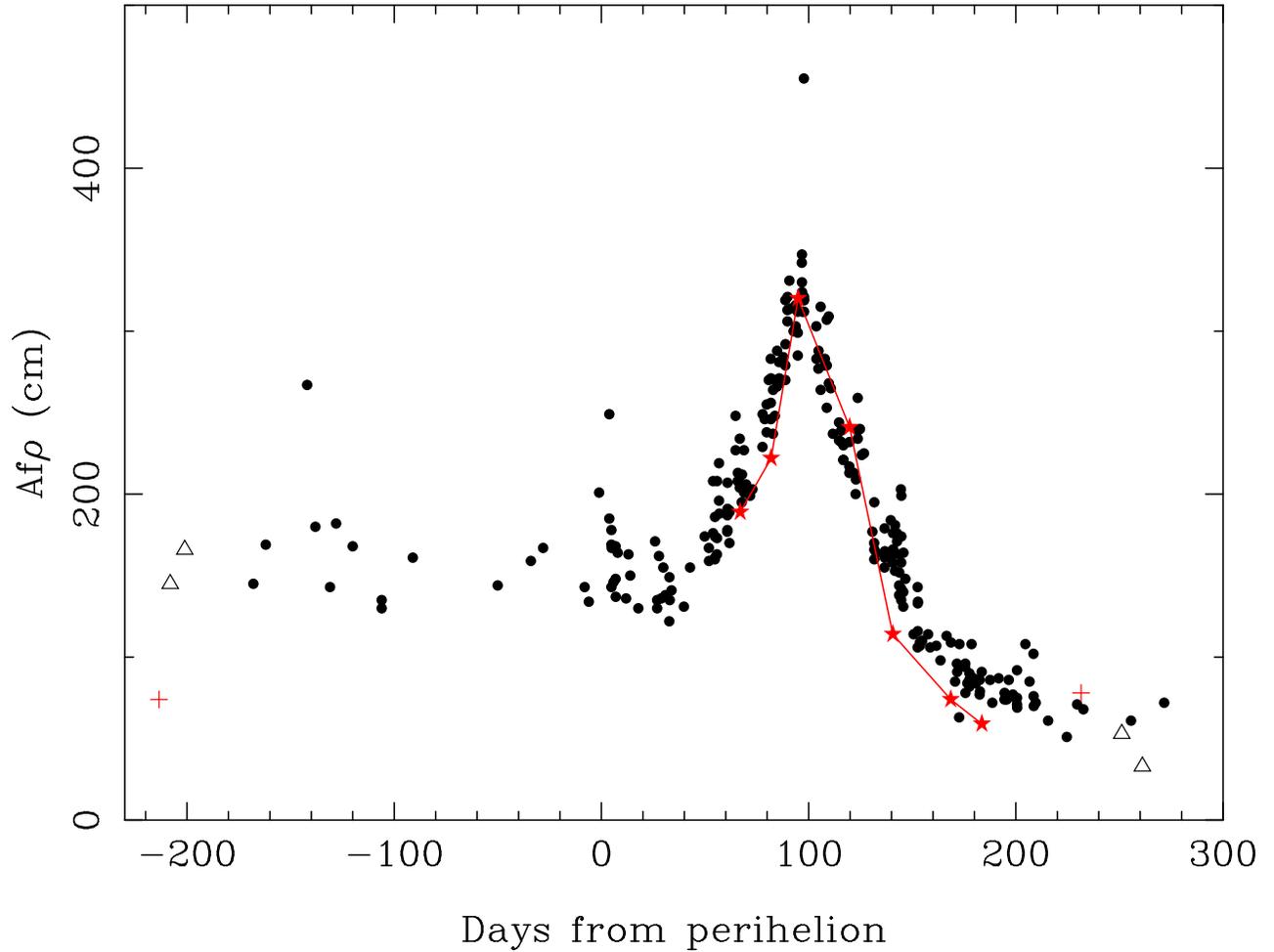}
\caption{Af$\rho$ versus days from perihelion. The filled circles correspond to the 2009 apparition data from {\it Cometas-Obs}, while the open triangles correspond to the 2002 apparition from the same group. The crosses (in red) correspond to the 2002 apparition data from Kiso (pre-perihelion) and CFHT (pos-perihelion) data. The star symbols joined by a solid line (in red) correspond to the 2009 apparition OSN data.}
\end{figure}
\clearpage

\begin{figure}
\includegraphics[angle=-90,scale=.80]{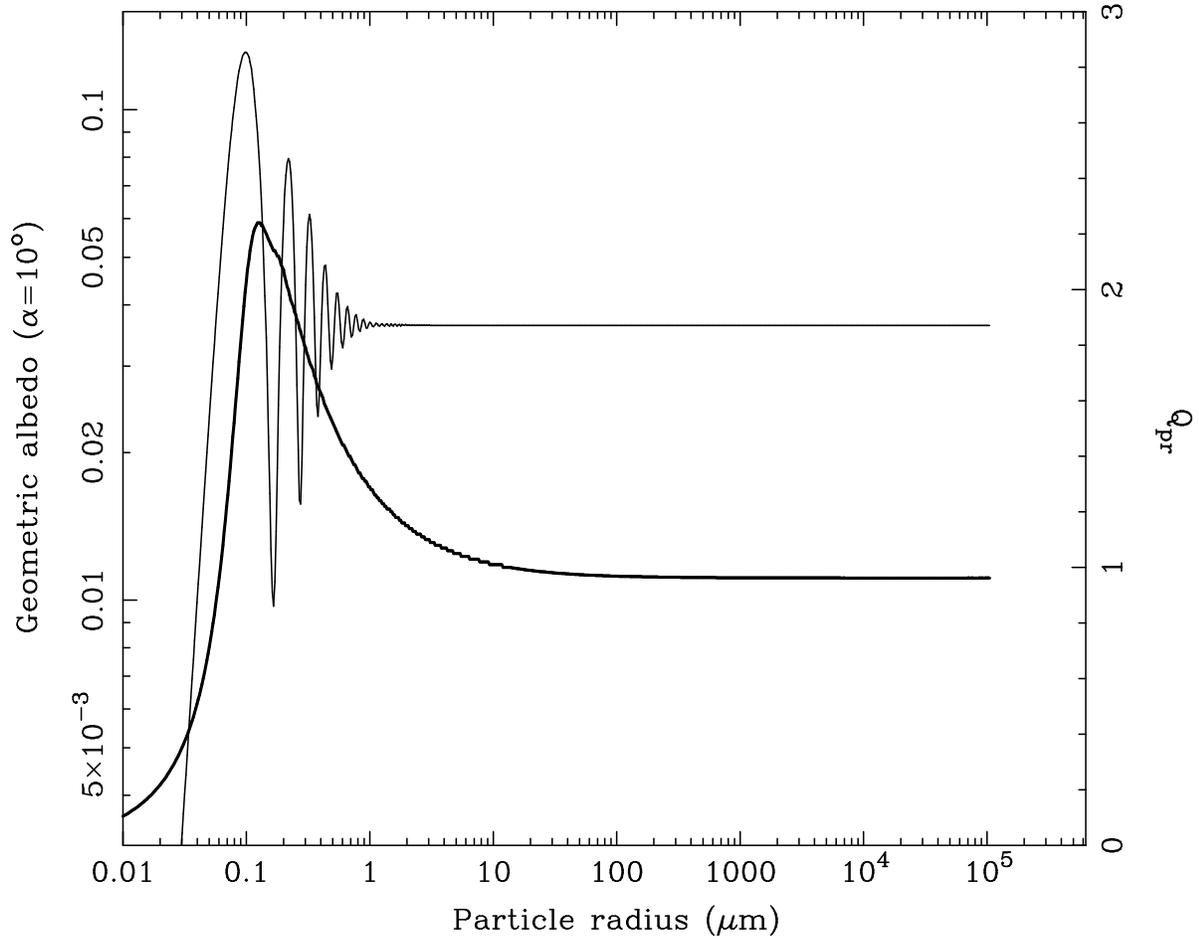}
\caption{Generalized geometric albedo (at 10$^\circ$ phase angle) versus particle radius for glassy carbon spheres (thin solid line, left axis). The thick line is the value of the radiation pressure coefficient, $Q_{pr}$, as a function of particle radius (right scale).}
\end{figure}
\clearpage

\begin{figure}
\includegraphics[angle=-90,scale=.80]{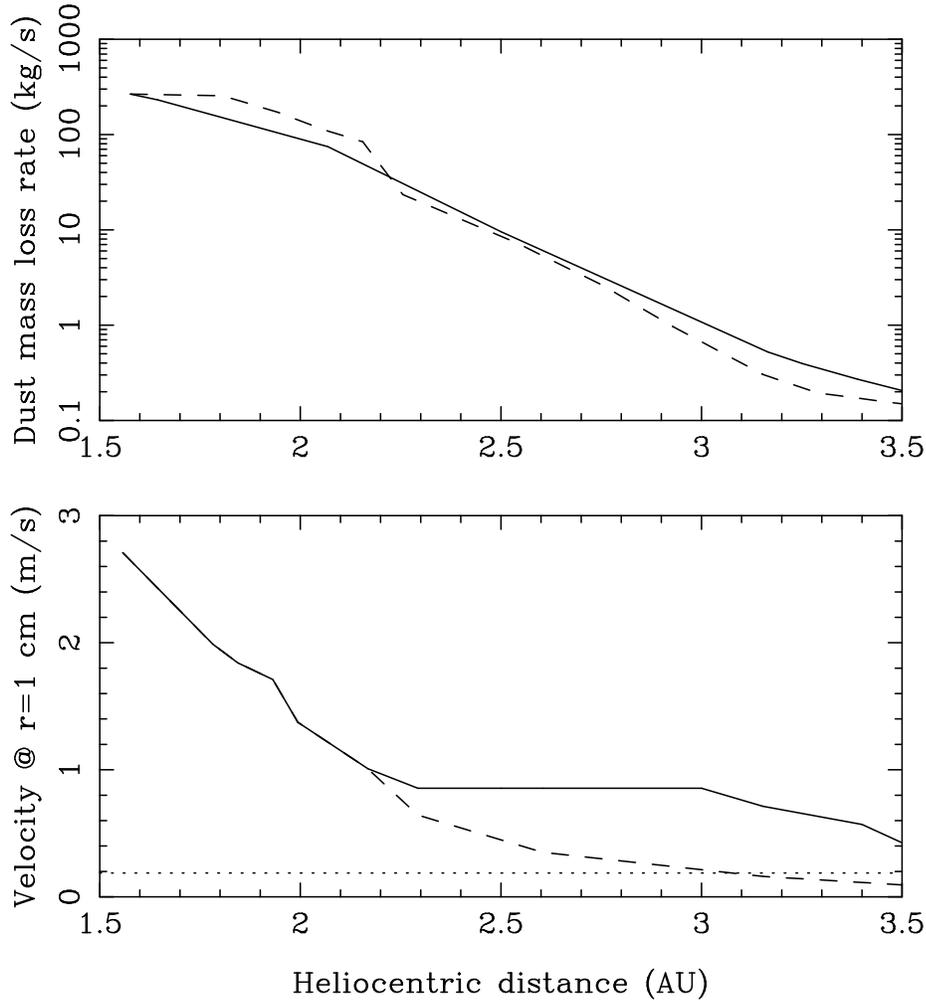}
\caption{Upper panel: Dust mass loss rate as a function of the heliocentric distance. Lower panel: ejection velocity of $r$=1 cm glassy carbon spheres as a function of heliocentric distance. In both panels, the solid line corresponds to pre-perihelion, and the dashed line to post-perihelion.}
\end{figure}
\clearpage

\begin{figure}
\includegraphics[angle=-90,scale=.75]{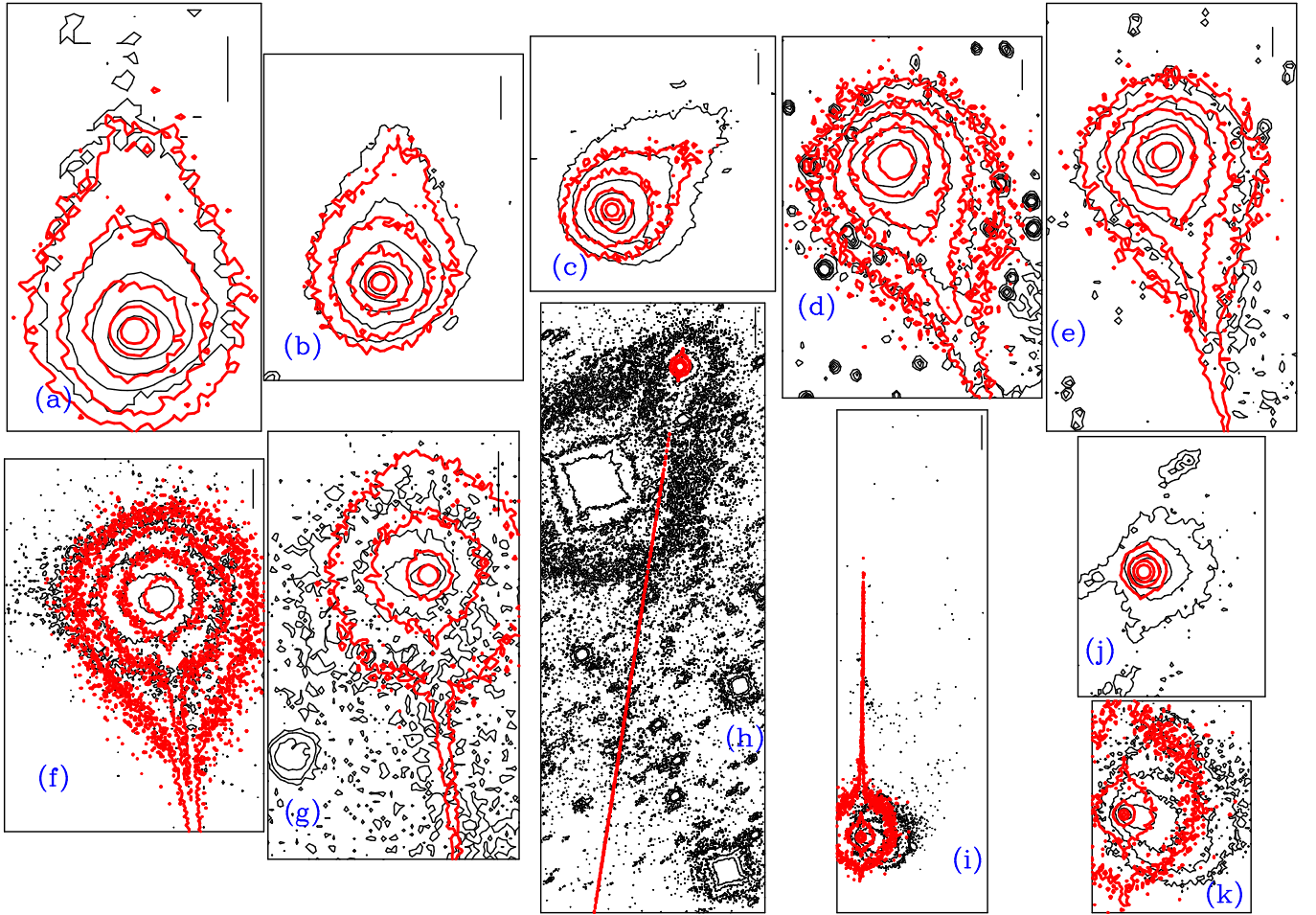}
\caption{Isotropic ejection model compared to the observations. The observation dates and layout corresponds to that shown in figure 1. In panels (a) to (c), the innermost isophotes are 3.2$\times$10$^{-13}$ solar disk intensity units; in panels (d) and (e), 8$\times$10$^{-14}$; in panel (f),  6$\times$10$^{-14}$; in panels (g) and (h),  1.2$\times$10$^{-13}$; in panel (i),  1.2$\times$10$^{-13}$; in panel (j),  4.8$\times$10$^{-14}$; and in panel (k), 2.4$\times$10$^{-13}$. Isophotes vary in factors of 2 between consecutive levels. In all panels, thin (black) contours correspond to the observations, and thick (red) contours correspond to the model. Vertical bars represent 20000 km in the images, except in (h), where it indicates 80000 km.}
\end{figure}
\clearpage

\begin{figure}
\includegraphics[angle=-90,scale=.80]{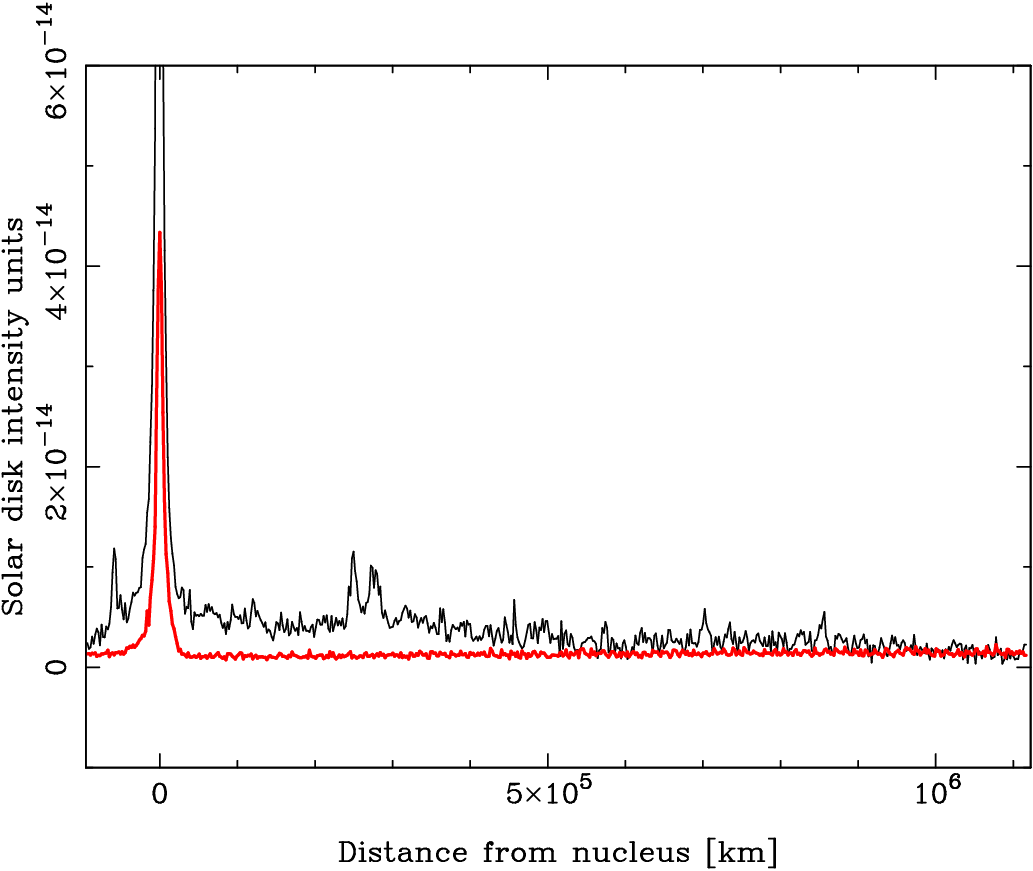}
\caption{Intensity scans along the trail in figure 7, panel (h). The thin (black) line correspond to the observation, while the thick (red) line is the model.}
\end{figure}
\clearpage

\begin{figure}
\includegraphics[angle=-90,scale=.75]{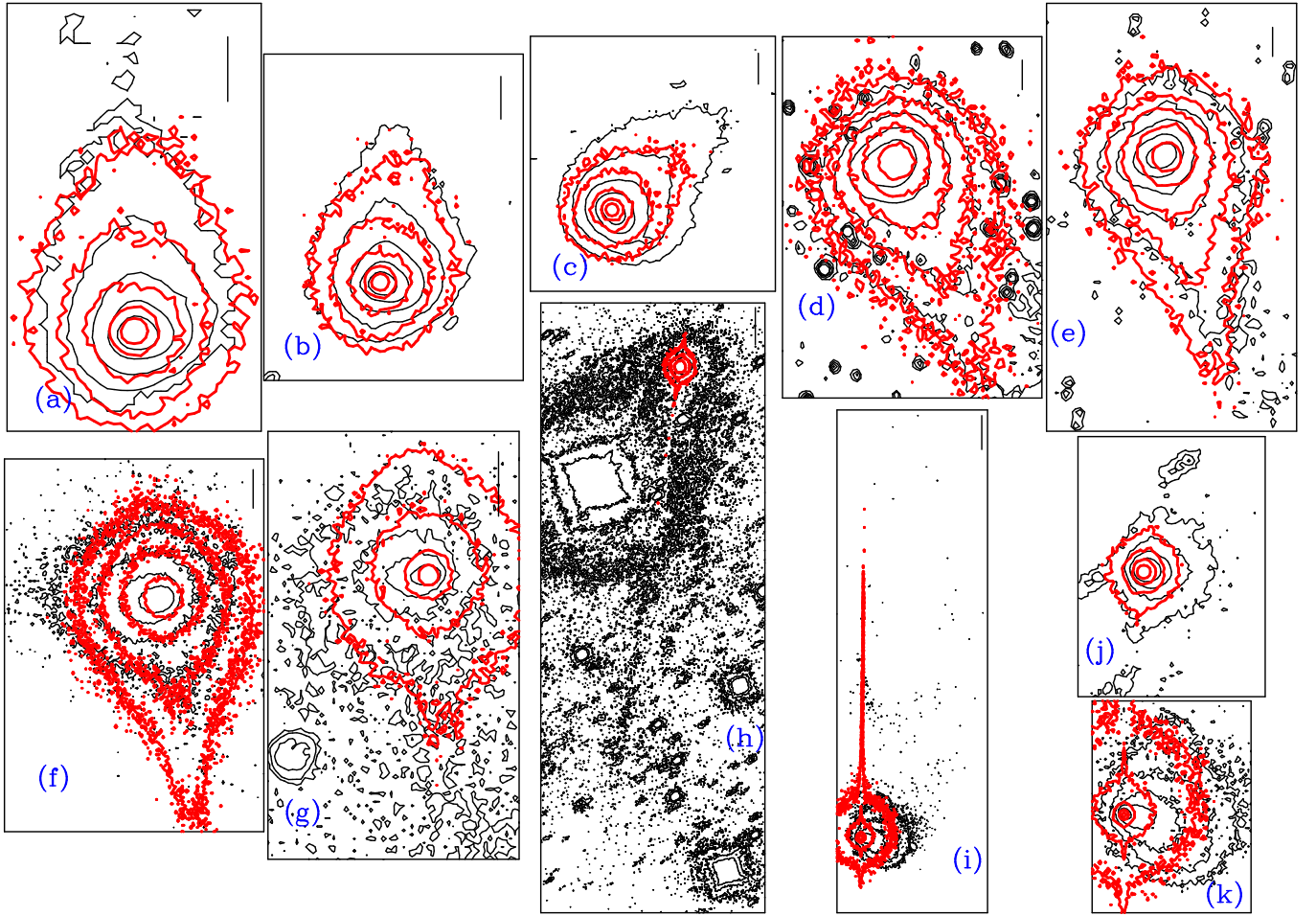}
\caption{Same as figure 7, but for the mass loss rate being multiplied by a factor of 15, and the ejection velocity by a factor of 5, for $r_h>$ 2.5 AU pre-perihelion, with respect to the values shown in figure 6. These factors have been imposed in order to obtain a better fit for the coma region in the 2002-05-12 image (compare panels (k) and (j) with Figure 7(k) and 7(j). Also, the neck-lines appear broader and closer to the observations than in figure 7 (compare panel (d) to (g) in both graphs), an effect produced by the substantially higher velocities at $r_h>$ 2.5 AU.}
\end{figure}
\clearpage

\begin{figure}
\includegraphics[angle=-90,scale=.80]{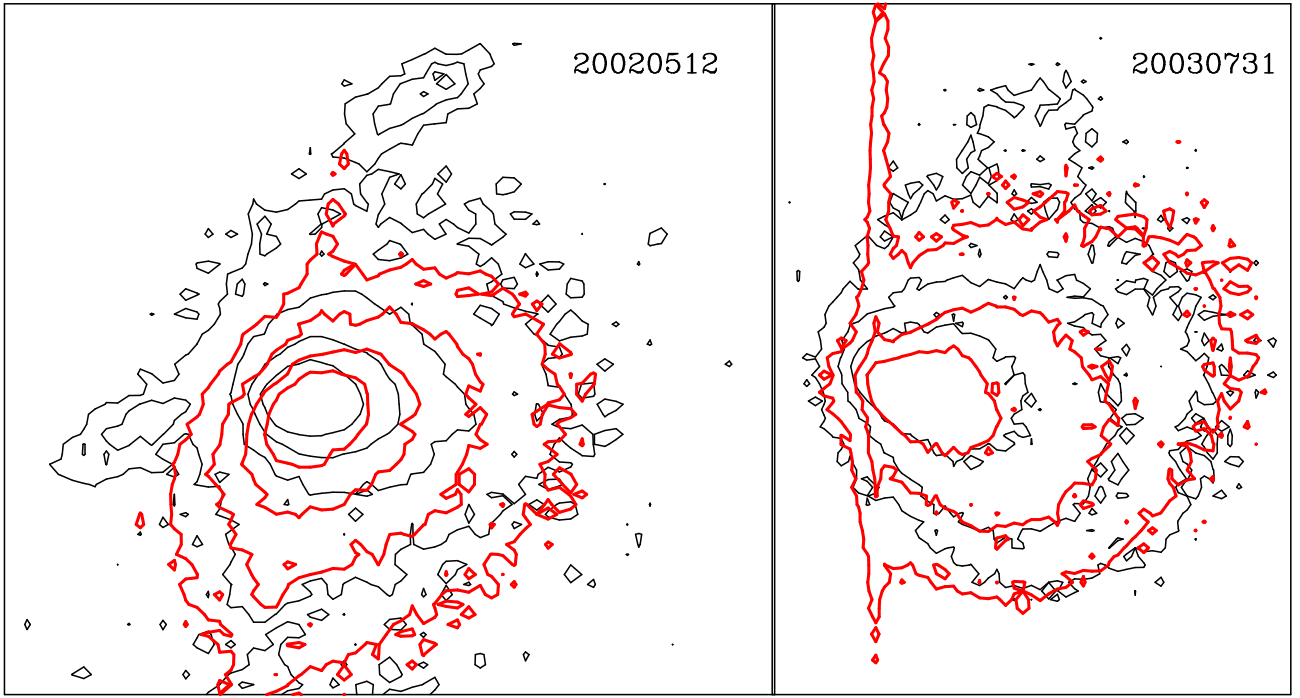}
\caption{Anisotropic ejection model applied to the 2002-05-12 and 2003-07-31 images (see Table I for the observational parameters and Table II for the model parameters). As in figure 9, the mass loss rate and the velocity have been increased at $r_h>$2.5 pre-perihelion. In this case, a factor of 30 has been applied to the mass loss rate, and the same factor of 5 in velocity.}
\end{figure}
\clearpage

\begin{figure}
\includegraphics[angle=-90,scale=.80]{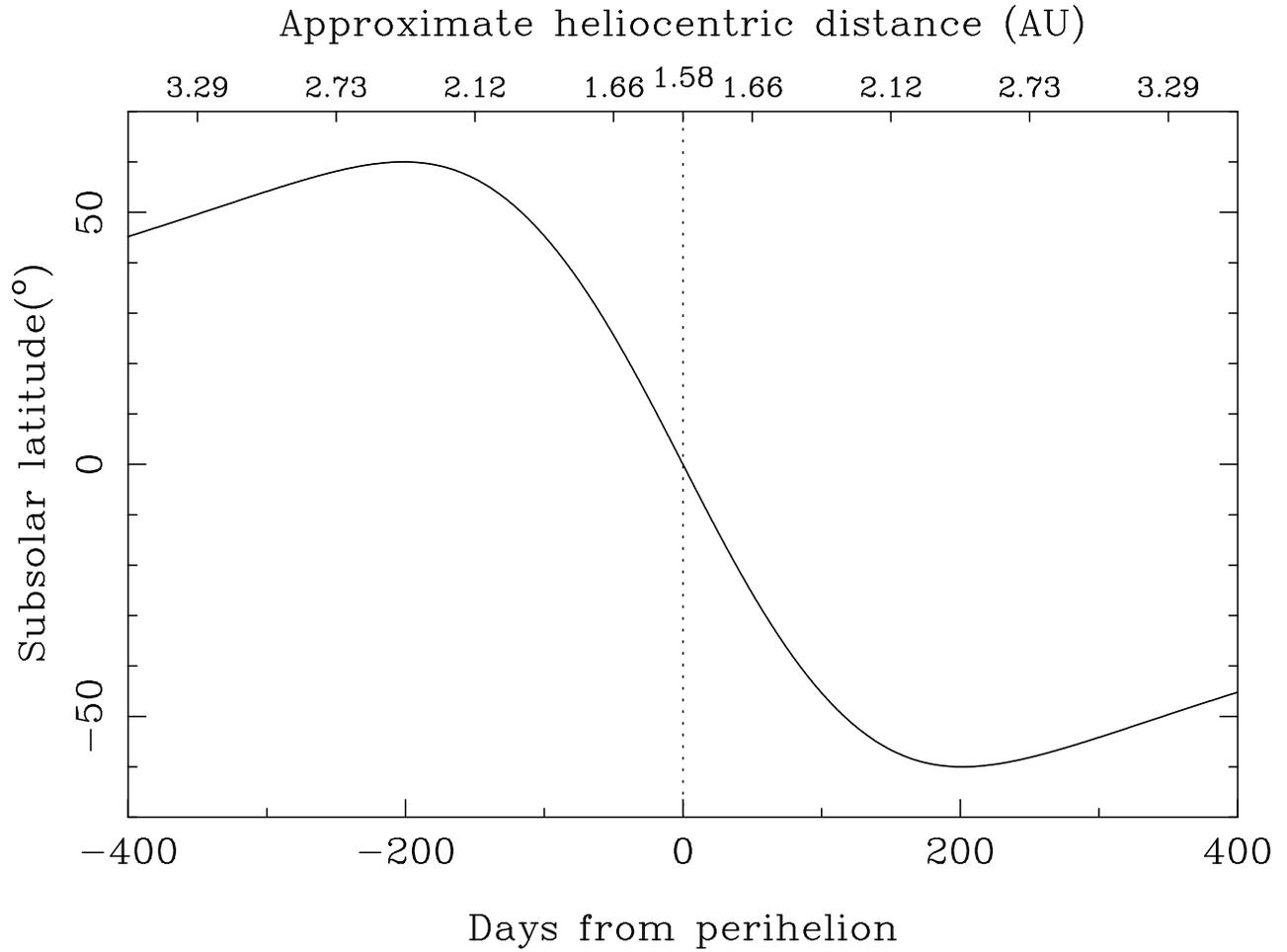}
\caption{Latitude of the subsolar point of 22P/Kopff nucleus as a function of time to perihelion, and heliocentric distance, for the rotational parameters $\Phi$=180$^\circ$, $I$=60$^\circ$ (see text).} 
\end{figure}
\clearpage

\begin{figure}
\includegraphics[angle=-90,scale=.80]{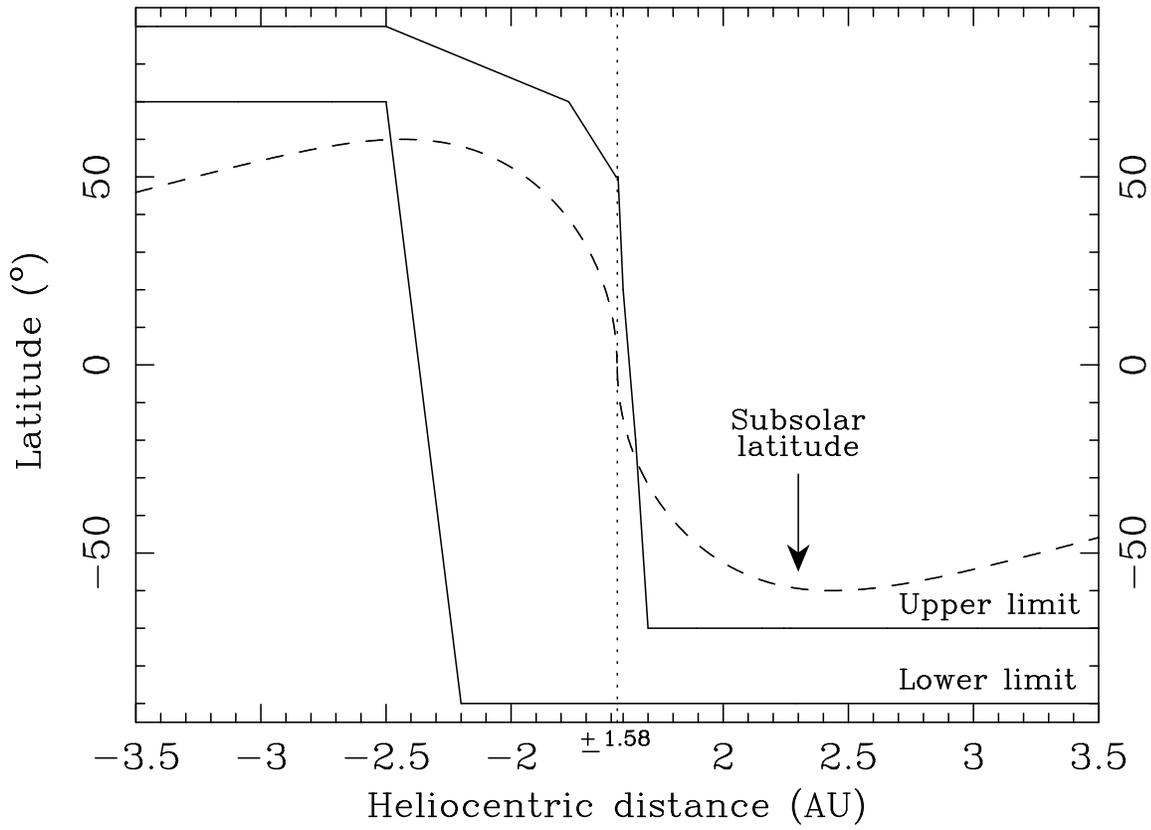}
\caption{Upper and lower latitude boundaries of the active area in the anisotropic dust ejection model, as a function of the heliocentric distance. Also displayed is the latitude of the subsolar point, which shows a similar behavior with time.}
\end{figure}
\clearpage

\begin{figure}
\includegraphics[angle=-90,scale=.75]{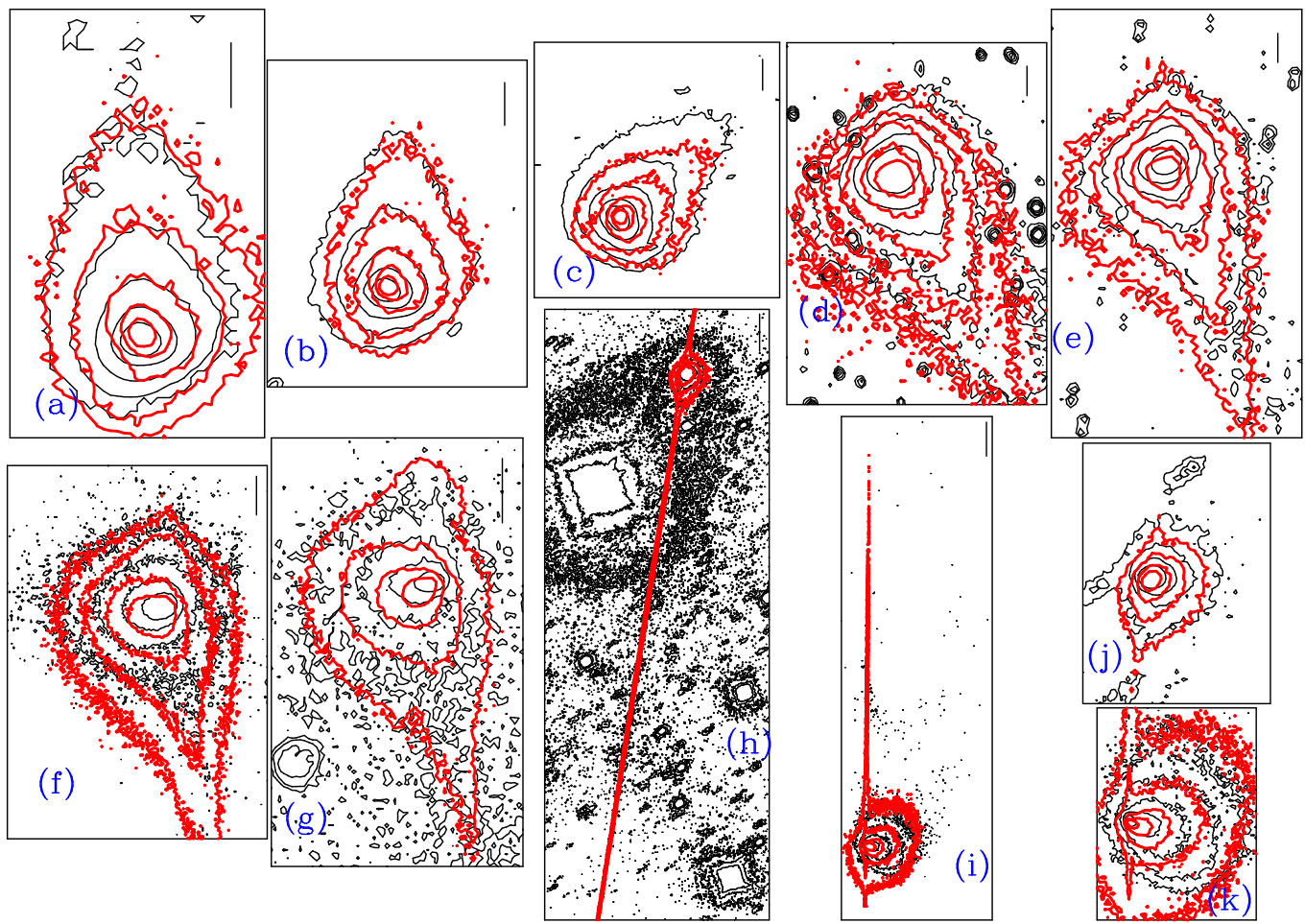}
\caption{Final version of the anisotropic model applied to all the available observations. For information on contours levels see figure 7. Thin (black) contours are the observations, and thick (red) contours, the model. The mass loss rates and ejection velocities are those of figure 6, but increased in factors of 20, and 5, respectively, at $r_h>$2.5 AU pre-perihelion. The vertical bars correspond to projected distances of 20000 km, except for panel (k), where it corresponds to 80000 km.}
\end{figure}

\clearpage

\begin{figure}
\includegraphics[angle=-90,scale=.80]{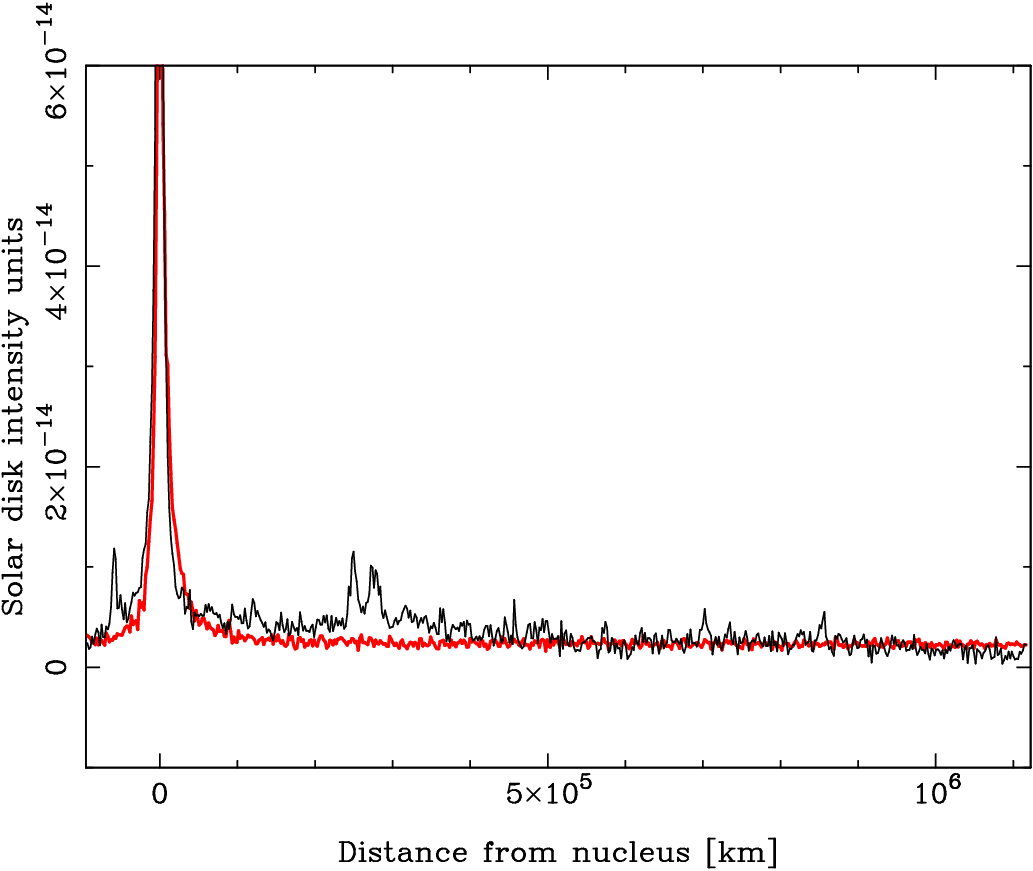}
\caption{Intensity scans along the trail in figure 13, panel (h). The thin (black) line correspond to the observation, while the thick (red) line is the model.}
\end{figure}

\clearpage
\begin{figure}
\includegraphics[angle=-90,scale=.80]{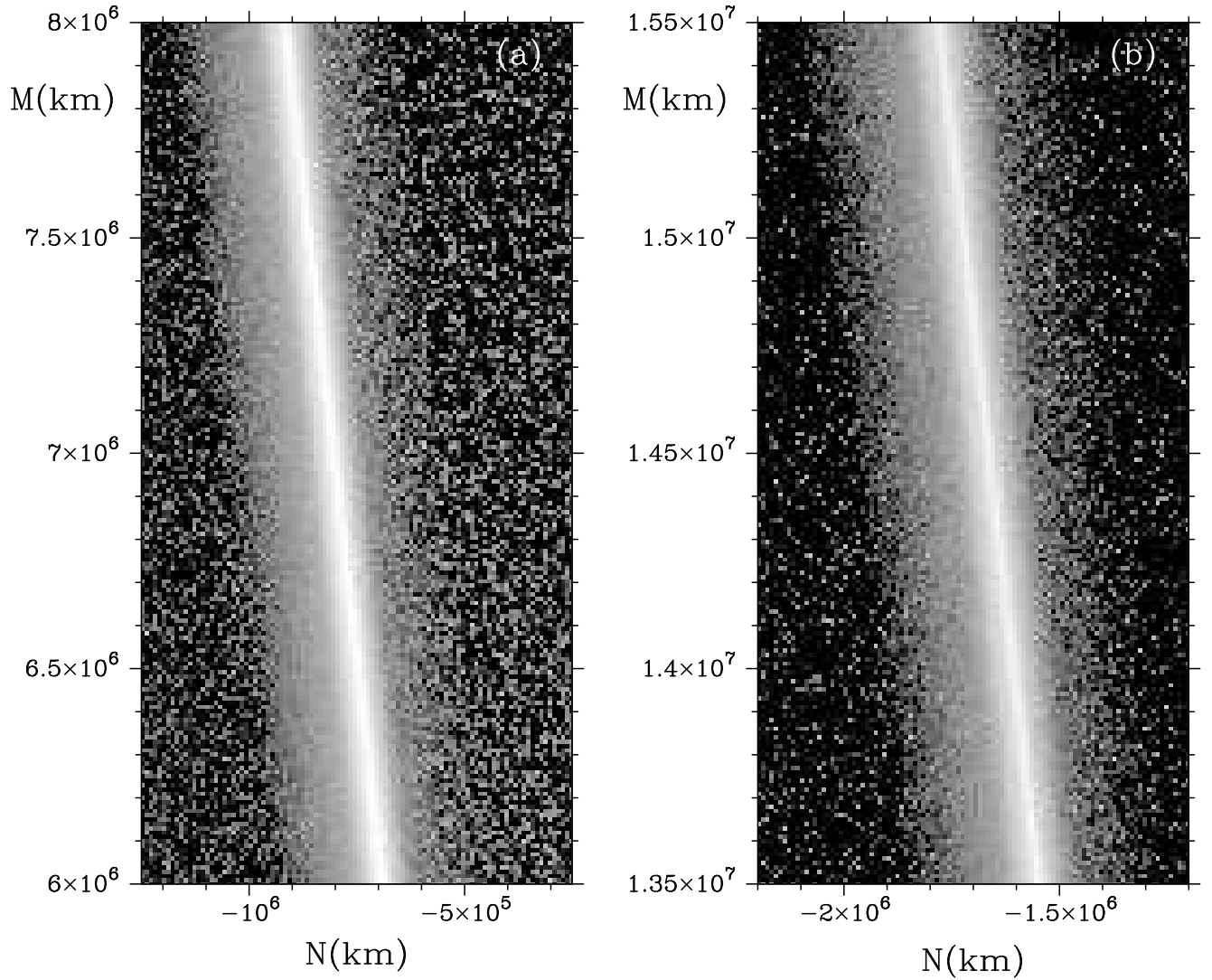}
\caption{Simulated trails at 12 $\mu$m wavelength on March 26, 1992, centered on regions of $\delta MA$=+0.5$^\circ$, panel (a), and $\delta MA$=+1.0$^\circ$, panel (b), as in the ISOCAM observations by \cite{Davies97}. The maximum intensity level in the images corresponds to 0.40 MJy sr$^{-1}$. Peak intensities are 0.40  MJy sr$^{-1}$, and 0.25 MJy sr$^{-1}$, at $\delta MA$=+0.5$^\circ$, and $\delta MA$=+1.0$^\circ$, respectively.}
\end{figure}

\clearpage

\begin{deluxetable}{crccrrr}
\tabletypesize{\scriptsize}
\rotate
\tablecaption{Log of the observations \label{tbl-1}}
\tablewidth{0pt}
\tablehead{
\colhead{Date}& \colhead{Days from} & \colhead{$r_h$} & \colhead{$\Delta$}  &  \colhead{Resolution} & \colhead{Phase} & \colhead{Af$\rho$} ($\rho$=10$^4$ km) \\
\colhead{[UT]} & \colhead{perihelion} & \colhead{[AU]} &  \colhead{[AU]} &    \colhead{[km pixel$^{-1}$]} & \colhead{Angle[$^\circ$]}  & \colhead{[cm]} \\
}
\startdata
2009 Jul 27 03:40 & 67.0 & 1.712 & 0.775 &  2068.5 & 19.6  &189  \\
2009 Aug 15 01:29 & 81.9 & 1.773 & 0.784 &  2098.9 & 10.7 &222 \\
2009 Aug 28 03:40 & 94.9 & 1.832 & 0.826 &  2204.6  & 3.9 &320 \\
2009 Sep 21 23:17 & 119.8 & 1.959 & 1.002 & 2669.0 & 12.4 & 241 \\
2009 Oct 12 20:10 & 140.6 & 2.075 & 1.242 & 3314.9 & 19.7 & 114 \\
2009 Nov 09 19:06 & 168.6 & 2.237 & 1.662 & 1109.0 & 24.1 &74 \\
2009 Nov 24 20:05 & 183.6 & 2.327 & 1.920 & 1281.1 & 24.6 & 59 \\ 
\tableline
2002 May 12 13:30 &--213.5   & 2.508 & 1.866 &  2029.6 & 20.8 & 74  \\
2003 Jul 31 14:17 &  231.5 & 2.615 & 2.508 &  673.0 & 22.7 & 78 \\
\enddata
\end{deluxetable}

\clearpage

\begin{deluxetable}{ll}
\tabletypesize{\scriptsize}
\tablecaption{Model physical parameters for anisotropic ejection models \label{tbl-2}}
\tablewidth{0pt}
\tablehead{
\colhead{Parameter} & \colhead{Value adopted/retrieved} 
}
\startdata
Grain density & 1000 kg m$^{-3}$ \\
Grain refractive index & m=1.88+0.71$i$ \\
Grain geometric albedo  & $p_v$=0.036 \\
Ejection velocity & $v(t,\beta)=v_1(t) \beta^{1/2}$, see Fig. 6 \\
Peak ejection velocity of 1-cm grains & 2.7 m s$^{-1}$ \\ 
Size distribution: r$_{min}$, r$_ {max}$ & 10$^{-4}$ cm, 1.4 cm \\
Size distribution: Power index & --3.1 \\
Peak dust mass loss rate (perihelion) & 260 kg s$^{-1}$ \\
Averaged dust mass loss rate per orbit & 40 kg s$^{-1}$ \\
Total dust mass ejected per orbit & 8$\times$10$^9$ kg \\
Pre-perihelion switch-on activity & r$_h\stackrel{>}{\sim}$3.5 AU \\
Nucleus rotation period & 12.3 h (Lowry and Weissmann, 2003) \\  
Argument of subsolar meridian at perihelion & $\Phi$=180$^\circ$ \\
Obliquity & $I$=60$^\circ$  \\
Active area location & Time-dependent, see Fig. 12 \\
Cone angle width & $\Delta\phi$=60$^\circ$ for $r_h>$2.5 AU pre-perihelion \\
                 & $\Delta\phi$=20$^\circ$ for $r_h>$1.95 AU post-perihelion \\
                 & $\Delta\phi$=40$^\circ$ otherwise \\
Isotropic emission percentage & 30\% \\  
\enddata
\end{deluxetable}

\clearpage

\begin{deluxetable}{lllll}
\tabletypesize{\scriptsize}
\tablecaption{Comparison of ISO and IRAS data with model results at 12$\mu$m wavelength\label{tbl-3}}
\tablewidth{0pt}
\tablehead{
\colhead{$\delta MA^\circ$} & \colhead{Measured} &  \colhead{Modeled} &  \colhead{Measured}  & \colhead{Modeled} \\
 & \colhead{brightness (MJy sr$^{-1}$)} &  \colhead{brightness (MJy sr$^{-1}$)} &  \colhead{FWHM (km)}  & \colhead{FWHM (km)} 
}
\startdata
+0.5 & 0.33$\pm$0.07 (ISO) & 0.40 & 48000$\pm$3000 & 26000 \\
+0.5 & 0.66$\pm$0.07 (IRAS) & 0.63 & 32000$\pm$16000 & 21000 \\
+1.0 & 0.26$\pm$0.07 (ISO) & 0.25 & 61000$\pm$3000 & 29000 \\
+1.0 & 1.35$\pm$0.11 (IRAS) & 0.32 & 32000$\pm$11000 & 31000 \\
\enddata
\end{deluxetable}


\begin{thebibliography}{}

\bibitem[A'Hearn et al.(1984)]{Ahearn84}A'Hearn, M.F., Schleicher, D.G., Millis, R.L., Feldman, P.D., and Thompson, D.T., 1984, \aj, 89, 579

\bibitem[Davies et al.(1997)]{Davies97}Davies, J.K., Sykes, M.V., Reach, W.T., et al., 1997, Icarus, 127, 251

\bibitem[Draine(2000)]{Draine00}Draine, B.T., 2000, in Light Scattering by Nonspherical Particles, M.I. Mishchenko, J.W. Hovenier, and L.D. Travis (eds.), Academic Press, pp. 131-145

\bibitem[Edoh(1983)]{Edoh83}Edoh, O., 1983, Optical Properties of Carbon from the Far Infrared to the Far Ultraviolet. Ph.D. dissertation, Univ. of Arizona

\bibitem[Finson \& Probstein(1968)]{Finson68} Finson, M., \& Probstein,
  R. 1968, \apj, 154, 327

\bibitem[Fulle(1989)]{Fulle89} Fulle, M., 1989, Astron. Astrophys.,
  217, 283

\bibitem[Fulle et al.(2010)]{Fulle10} Fulle, M., Colangeli, L.,
  Agarwal, J., et al. 2010, Astron. Astrophys., 522, 63

\bibitem[Groussin et al.(2009)]{Groussin09}Groussin, O., Lamy, P., Toth, I., et al., 2009, Icarus, 199, 568

\bibitem[Hanner et al.(1997)]{Hanner97}Hanner, M.S., Gehrz, R.D., Harker, D.E., et al., 1997, Earth, Moon, \& Planets, 79, 247

\bibitem[Ishiguro et al.(2007)]{Ishiguro07}Ishiguro, M., Sarugaku, Y., Ueno, M., Miura, N., Usui, F. Chun, M.-Y., and Kwon, S. M., 2007, Icarus, 189, 169

\bibitem[K\c{e}pi\'nski(1958)]{Kepinski58}K\c{e}pi\'nski, F. 1958, Acta Astronomica, 8, 193

\bibitem[K\c{e}pi\'nski(1963)]{Kepinski63}K\c{e}pi\'nski, F. 1963, Acta Astronomica, 13, 87


\bibitem[Kimura and Liu(1977)]{Kimura77}Kimura, H., \& Liu, C., 1977, Chin. Astron., 1, 235

\bibitem[Kimura et al.(2003)]{Kimura03}Kimura, H., Kolokolova, and L., Mann, I., 2003, Astron. Astrophys., 407, L5

\bibitem[Kolokolova et al.(2004)]{Kolokolova04}Kolokolova, L., Hanner, M.S., Levasseur-Regourd, A.-Ch., and Gustafson, B.A.S., 2004, in Comets II, M. C. Festou, H. U. Keller, and H. A. Weaver (eds.), University of Arizona Press, Tucson, 745 pp., p.577-604

\bibitem[Kraemer et al.(2005)]{Kraemer05}Kraemer, K.E., Lisse, C.M., Price, S.D., et al., 2005, \aj, 130, 2363

\bibitem[Lamy et al.(2002)]{Lamy02}Lamy, P.L., Toth, I., Jorda, L., Groussin, O., A'Hearn, M.F., and Weaver, H.A., 2002, Icarus, 156, 442

\bibitem[Lowry and Fitzsimmons(2001)]{Lowry01}Lowry, S.C., \& Fitzsimmons, A., 2001, Astron. Astrophys., 365, 204

\bibitem[Lowry and Weissmann(2003)]{Lowry03}Lowry, S.C., \& Weissman, P.R., 2003, Icarus, 164, 492

\bibitem[Meech and Jewitt(1987)]{Meech87}Meech, K., \& Jewitt, D., 1987, Astron. Astrophys. 187, 585

\bibitem[Mishchenko et al.(2000)]{Mishchenko00}Mishchenko, M.I., Travis, L.D., and Macke, A., 2000, in Light Scattering by Nonspherical Particles, M.I. Mishchenko, J.W. Hovenier, and L.D. Travis (eds.), Academic Press, pp. 131-145

\bibitem[Monet et al.(2003)]{Monet03} Monet, D.G., Levine, S.E.,  
Canzian, B. et al. 2003, \aj, 125, 984

\bibitem[Moreno et al.(2007)]{Moreno07}Moreno, F.; Mu\~noz, O., Guirado, D., and Vilaplana, R., 2007, J. Quant. Spectrosc. Rad. Transfer, 106, 348

\bibitem[Moreno(2009)]{Moreno09} Moreno, F. 2009, \apjs, 183, 33

\bibitem[Moreno et al.(2011)]{Moreno11}Moreno, F., Lara, L.M.; Licandro, J., et al., 2011, \apj, 738, L16

\bibitem[Nelder and Mead(1965)]{Nelder65}Nelder, J.A., Mead, R., 1965, Computer Journal, 7, 308
 
\bibitem[Press et al.(1992)]{Press92}Press, W.H., Teukolsky, S.A., Vetterling, W.T., and Flannery, B.P., 1992, in Numerical Recipes in FORTRAN, Cambridge University Press, Cambridge, pp. 402-406

\bibitem[Rickman et al.(1984)]{Rickman87}Rickman, H., Sitarski, G., and Todorovic-Juchniewicz, B., 1987, Astron. Astrophys., 188, 206

\bibitem[Sekanina(1981)]{Sekanina81}Sekanina, Z., 1981, Ann. Rev. Earth and Planet. Sci., 9, 113

\bibitem[Sekanina(1984)]{Sekanina84}Sekanina, Z., 1984, \aj 89, 1573

\bibitem[Sosa and Fern\'andez(2009)]{Sosa09}Sosa, A., \& Fern\'andez, J.A., 2009, \mnras, 393, 192


\bibitem[Sykes and Walker(1992)]{Sykes92}Sykes, M.V., \& Walker, R.G., 1992, Icarus, 95, 180

\bibitem[Tancredi et al.(2000)]{Tancredi00}Tancredi, G., Fern\'andez, J.A.,  Rickman, H., \& Licandro, J., 2000, Astron. Astrophys., 146, 73	

\bibitem[Yeomans(1974)]{Yeomans74}Yeomans, D.K., 1974, \pasp, 86, 125

\bibitem[Zubko(2011)]{Zubko11}Zubko, E., 2011, Light scattering by irregularly shaped particles with sizes comparable to the wavelength, in Light Scattering Reviews, vol. 6: Light Scattering and Remote Sensing of Atmosphere and Surface, A.A. Kokhanovsky (ed.), Springer-Praxis, Berlin, pp. 39-74

\end{thebibliography}
\end{document}